\documentclass[11pt,a4paper]{article}
\pdfoutput=1

\usepackage{jcappub}
\usepackage{amssymb,amsmath}
\usepackage{graphicx}
\usepackage{epsfig}

\newcommand{\beq}{\begin{equation}}
\newcommand{\eeq}{\end{equation}}
\newcommand{\bea}{\begin{eqnarray}}
\newcommand{\eea}{\end{eqnarray}}
\newcommand{\vc}[1]{{\textbf{#1}}}
\newcommand{\mc}[1]{\mathcal{#1}}

\title{Effective long wavelength scalar dynamics in de Sitter}

\author{Ian Moss and}
\author{Gerasimos Rigopoulos}
\affiliation{School of Mathematics and Statistics, 
	Herschel Building, Newcastle University, 
	Newcastle upon Tyne, NE1 7RU, UK}

\abstract{ 
We discuss the effective infrared theory governing a light scalar's long wavelength dynamics in de Sitter spacetime. 
We show how the separation 
of scales around the physical curvature radius $k/a \sim H$ can be performed consistently with a window function 
and how short wavelengths can 
be integrated out in the Schwinger-Keldysh path integral formalism. At leading order, and for time scales $\Delta t \gg H^{-1}$, this results in the well-known Starobinsky stochastic 
evolution. However, our approach allows for the computation of quantum UV corrections, generating an effective potential on which the stochastic dynamics takes place. The long wavelength stochastic dynamical equations are now second order in time, incorporating temporal scales $\Delta t \sim H^{-1}$ and resulting in a Kramers equation for the probability distribution - more precisely the Wigner function - in contrast to the more usual Fokker-Planck equation. This feature allows us to non-perturbatively evaluate, within the stochastic formalism, not only expectation values of field correlators, but also the stress-energy tensor of $\phi$. 
}

\begin{document}
\maketitle

\section{Introduction}
The quantum dynamics of a scalar field $\phi$ in de Sitter spacetime geometry is among the simplest examples of a quantum field theory in a 
curved spacetime. It is also highly relevant physically as it forms the basis of understanding quantum fluctuations during inflation. 
Nevertheless, it is still far from being completely understood. The main issues revolve 
around the behaviour of the field's correlations on long wavelengths and over long times. These correlations are usually referred to 
as exhibiting infrared (IR) and secular divergences, first recognized in \cite{Ford:1977in}. Similar divergences are found in the theory of inflationary 
perturbations when curvature correlations are computed beyond the leading order. A comprehensive review of these issues and extensive references to the relevant works can be found in 
\cite{Seery:2010kh} - see also \cite{Woodard:2014jba} for a concise 
view on related issues. Given the central position of this physical model for our current conception the early universe, a complete 
elucidation of its IR physics seems pertinent and could shed further light into the theory of inflationary perturbations where also gravitational fluctuations are relevant.                   

A major conceptual advance in understanding the IR behaviour of $\phi$ in (quasi-)de Sitter was the realization by Starobinsky 
that, owing to the non-oscillatory behavior of long wavelength modes, the quantum dynamics of $\phi$ on super-Hubble scales 
reduces to that of a classical field in the presence of a stochastic force \cite{Starobinsky:1986fx, Starobinsky:1994bd}. Although physically appealing, the foundations of this idea remained somewhat obscure. A significant step to clarify the relationship of this picture to the underlying Quantum Field Theory (QFT) was taken by Woodard \cite{Woodard:2005cw} who demonstrated that stochastic inflation reproduces the 
leading order scale factor logarithms appearing in perturbative QFT computations of coincident expectation values. Unlike the QFT framework however, the stochastic picture allows one to re-sum the uncontrolled perturbative expansion and obtain finite answers. 

The comparison of the stochastic formalism to the full QFT computation was further studied in \cite{vanderMeulen:2007ah, Finelli:2008zg, Garbrecht:2013coa, Garbrecht:2014dca, Onemli:2015pma} and its results for the final equilibrium distribution for $\phi$ agree, where comparisons can be made, with other computational methods such as Euclidean computations or use of the non-perturbative renormalization group \cite{Rajaraman:2010xd, Beneke:2012kn, Guilleux:2015pma, Gautier:2015pca, Nacir:2016fzi}. In \cite{Garbrecht:2013coa, Garbrecht:2014dca} it was shown 
that the full QFT and the stochastic descriptions of the field's dynamics coincide exactly in their diagrammatic expansions in the IR 
and hence all expectation values of powers of $\phi$ computed in the two frameworks should also coincide in the IR. The appropriateness of the stochastic framework for the describing the IR effective theory in inflation was recently also advocated in \cite{Burgess:2015ajz}.

Many of the above works include (favorable) comparisons of computations in the stochastic formalism to corresponding computations using the full QFT machinery. Given the evidence in favour of its validity, the stochastic picture has been extensively used by cosmologists studying inflation \cite{Nambu:1988je, Kandrup:1988sc, Salopek:1990re, Mollerach:1990zf, Habib:1992ci, GarciaBellido:1994vz, Liguori:2004fa, Rigopoulos:2004gr, Martin:2005ir, Prokopec:2007ak, Adshead:2008gk, Riotto:2008mv, Finelli:2010sh, Enqvist:2011pt, Kuhnel:2010pp, Riotto:2011sf, Weenink:2011dd,Martin:2011ib, Hwang:2012mf, Rigopoulos:2013exa, Levasseur:2013tja, Lazzari:2013boa, Burgess:2014eoa, Vennin:2015hra, Vennin:2016wnk}. Nevertheless, its basis on the full quantum theory has remained somewhat obscure since, apart from Starobinsky's original argument which was limited to the leading order, no first principles derivation has been put forward. It has therefore been considered by many field theorists as a heuristic tool at best whose results would need to be checked against more refined computations. Furthermore some misunderstandings exist in the literature regarding the nature of the stochastic force which, unlike more common system-environment computations, exists even for a free theory, being the result of a time dependent system-environment split.\footnote{For a description of stochastic dynamics due to a standard system-environment interaction in an inflationary setting see \cite{Boyanovsky:2015tba, Boyanovsky:2015jen}}  

In this work we further elucidate the framework of the stochastic 
theory of $\phi$ during inflation by explicitly computing the IR Effective Field Theory which results after integrating out short wavelength fluctuations. We use the appropriate Schwinger-Keldysh amphichronous path integral, extending the results of \cite{Rigopoulos:2016oko} beyond the leading order. By ``short wavelength'' we mean fluctuations with physical wavelengths below the curvature scale of de Sitter, $k/a > H$. An initial formulation along similar lines was given by \cite{Morikawa:1989xz} and a related approach was discussed in \cite{Levasseur:2013ffa}. The IR effective theory we derive here is fully renormalized in the UV and consists of a generalization of Starobinsky's stochastic inflation, extended to include quantum corrections from sub-Hubble quantum fluctuations. As expected, the latter shift the potential on which the stochastic evolution takes place. Once these UV quantum fluctuations are properly accounted for, the theory is IR safe with IR resummations afforded by the stochastic dynamics. Our  formulation also accounts correctly for processes that operate on timescales of order $\Delta t \gtrsim H^{-1}$, unlike Starobinsky's original formulation which corresponds to the long time/strong friction limit $\Delta t \gg H^{-1}$.  

On the technical side, our methodology is somewhat similar to a Wilsonian integrating out of fast/short wavelength dynamics to obtain the relevant slow/long wavelength theory, adopted to the peculiarities of a time dependent split between the two sectors. We expect that a generalization of this approach will be possible and useful in obtaining the IR effective theory for more complicated field theories in (quasi-)de Sitter spacetime, for example including gravitational fluctuations - see \cite{Tsamis:2005hd} for a first attempt in this direction. 
        
\section{Long wavelength stochastic dynamics}
In this section we systematically derive the long wavelength effective field theory, obtaining Starobinsky's stochastic inflation 
along with its leading order quantum corrections which shift the IR potential from its bare form.\footnote{There is also an additional, subdominant, component of the stochastic noise, see (\ref{extra stoch}).} After briefly 
reviewing the CTP path integral we proceed to separate scales through the use of a window function. We then 
explicitly integrate out the short wavelength sector using the Schwinger-Keldysh formalism, obtaining the long 
wavelength effective action in a gradient expansion which includes the leading order quantum corrections. Our 
discussion is fairly general and relies on the infrared enhancement of the propagator on long separations relative 
to the Minkowski result. We then specialize in de Sitter, deriving the IR effective potential for a light scalar with a 
quartic bare interaction potential.             

\subsection{The CTP formalism}
\label{ctpformalism}

The real time evolution of the scalar field from given initial conditions, 
without reference to a future asymptotic state, is described by the so-called Schwinger-Keldysh formalism employing the 
amphichronous or closed-time-path (CTP) path integral \cite{Calzetta:2008iqa, Altland:2006si}. This replaces the
scalar field with two scalar fields $\phi_a$ ($a=1,2$) and generating functional
\beq  \label{CTP}
Z\left[\boldsymbol{J}\right] = \int D\boldsymbol{\phi}\, \exp \frac{i}{2} 
\int\left( \boldsymbol{\phi}^\dagger\boldsymbol{D}^{-1}\boldsymbol{\phi}
-2\overline{V}_I(\boldsymbol{\phi})+\boldsymbol{J}^\dagger\boldsymbol{\phi}\right)d\mu\,, 
\eeq
where 
\beq
\boldsymbol{\phi}=\left(\begin{matrix}
	\phi_1\\\phi_2	
\end{matrix}\right)\,,\quad 
\boldsymbol{\phi}^\dagger=(\phi_1,\phi_2)\mathbf{C}\,,\quad 
\overline{V}_I(\boldsymbol{\phi})=V_I(\phi_1)-V_I(\phi_2)\,,\quad 
\mathbf{D}^{-1}(x,x')=(\nabla^2-m^2){\bf 1}\delta(x,x'),
\eeq
$d\mu =  d^4x\sqrt{-g}$ is the volume measure and $\int\delta(x,x')d\mu=1$.
The adjoint operation $\dagger$ is defined by a metric on field space $\mathbf{C}$ which can be used to raise the `$a$' index. In the above 
Schwinger basis it has components $C^{ab}={\rm diag}(1,-1)$. The path integral representation (\ref{CTP}) of the generating functional corresponds to 
\beq\label{CTP-U}
Z[J_1,J_2] = \lim_{T \to \infty}  {\mathrm Tr} \left[U_{J_2}(0,T)U_{J_1}(T,0)\rho(0)\right]
\eeq
which describes evolution forward in time from $t=0$ to $t=T$ with an initial density matrix $\rho(0)$ and evolution operator $U_{J_1}$ in the presence of an external current $J_1$, and then evolution backwards in time in the presence of a different current $J_2$. The boundary conditions 
implied in (\ref{CTP}) involve the initial density matrix $\rho$ (or some given state in the past, e.g. the vacuum) and that $\phi_1=\phi_2$ at some 
point in the future after any 
possible time of interest. The latter condition results from inserting a complete set of field eigenstates at $T$ in (\ref{CTP-U}). This type of 
generating functional, and the corresponding doubling of fields in (\ref{CTP}), is necessary for obtaining the correct causal dynamical 
evolution for expectation values of field operators \cite{Calzetta:2008iqa}. The boundary conditions implicit in (\ref{CTP}) determine the
choice of the Green function matrix which satisfies,
\beq
(\nabla^2-m^2)\,\mathbf{D}_a{}^b(x,x')=\delta_a{}^b\delta(x,x').
\eeq
In the Schwinger $\phi_{1,2}$ basis with both field indices lowered,
\beq
\mathbf{D}_{ab}(x,x')=\left(\begin{matrix}
	D_{11}(x,x')&	D_{12}(x,x')\\	D_{21}(x,x') &	D_{22}(x,x')	
\end{matrix}\right)= -i\left(\begin{matrix}
\langle T \phi(x)\phi(x')\rangle & \langle\phi(x')\phi(x)\rangle\\
\langle\phi(x)\phi(x')\rangle & \langle\tilde{T}\phi(x)\phi(x')\rangle	
\end{matrix}\right)\,.\label{gfsb}
\eeq

Instead of using $\phi_1$ and $\phi_2$, it's possible to change the field description via 
\beq
\boldsymbol{\phi}' = \mathbf{U}\boldsymbol{\phi} \,.
\eeq
For example, it is often helpful to employ the ``classical'' and ``quantum'' fields of the Keldysh basis
\beq
\phi=\frac{\phi_1+\phi_2}{2}\,,\quad {\phi_q}=\phi_1-\phi_2\,,
\eeq
with transformation matrix  
\beq
\mathbf{U}_a{}^b=\left(\begin{matrix}
	\frac{1}{2}& 	\frac{1}{2} \\1 & -1
\end{matrix}\right)\,.
\eeq
In this basis,
\beq
{\bf C}^{\prime ab}=\left(\begin{matrix}0&1\\1&0\end{matrix}\right)
\eeq
and accordingly
\beq
\mathbf{D}^{\prime-1ab}(x,x')= \left(\begin{matrix}
	0& (\nabla^2-m^2) \\ (\nabla^2-m^2) & 0	
\end{matrix}\right)\delta(x,x')\,.
\eeq 
The Green function (\ref{gfsb}) in the Keldysh basis is 
\beq\label{Keldysh-prop}
\mathbf{D}'_{ab}=
\left(\begin{matrix}
-i D_K(x,x')&	D_{\rm R}(x,x')\\	D_{\rm A}(x,x') & 0	
\end{matrix}\right)
= -i \left(\begin{matrix}
	\frac{1}{2}\langle \{\phi(x),\phi(x')\} \rangle & \langle[\phi(x),\phi(x')]\rangle 
	\theta(t-t')\\ \langle[\phi(x'),\phi(x)]\rangle \theta(t'-t)& 0
\end{matrix}\right)\,,
\eeq
where $D_{R(A)}$ is the retarded (advanced) Green function and $D_K$ is the 
Keldysh (or Hadamard) component of the propagator. The coupling to currents reads in the Keldysh basis 
\beq
\mathbf{J}^\dagger\boldsymbol{\phi} = J{\phi_q}+J_q\phi \,,
\eeq
while 
\beq\label{Vbar}
\overline{V}_I= \frac{\partial V_I}{\partial\phi}{\phi_q} +  \!\sum\limits_{m=1}^\infty\!\frac{V_I^{{(2m+1)}}}{2^m\left(2m+1\right)!}\,{\phi_q}^{2m+1}\,.
\eeq
It is worth noting that classical physics is obtained in the limit ${\phi_q}\rightarrow 0$. Keeping $\mc{O}({\phi_q}^2)$ terms corresponds to making the dynamics stochastic (and possibly incorporating stochastic initial conditions).

Perturbation theory is built similarly to the more usual in-out path integral, using the matrix $\mathbf{D}$ as the propagator, with one notable difference. Setting $V_I$ and $\vc{J}$ to zero we immediately have
\beq\label{Z0}
Z_0\left[\boldsymbol{0}\right] = \int D\boldsymbol{\phi}\, \exp \frac{i}{2} 
\int \boldsymbol{\phi}^\dagger\boldsymbol{D}^{-1}\boldsymbol{\phi}
\,\,d\mu= 1 
\eeq  
since it is simply $U_0U_0^\dagger=1$ for unitary evolution. This implies that we must have
\beq\label{Z0-2}
{\rm Tr}\ln \left(\begin{matrix}
-i D_K(x,x')&	D_{\rm R}(x,x')\\	D_{\rm A}(x,x') & 0	
\end{matrix}\right) =0
\eeq
in any representation of the propagator which may be used to define the above trace. Unlike the in-out vacuum-to-vacuum amplitude used in scattering computations, $Z_0[\mathbf{0}]=1$ and there is no need to factor out vacuum bubbles as they are automatically zero. Indeed, unitarity
of the time evolution operator $U_{J}(T,0)$ implies a stronger result for the full, interacting theory,
\beq
\left.Z[{\bf J}]\right|_{J_q=0}=1,
\eeq
and the effective action constructed from 1PI Feynman diagrams will also vanish when $\phi_q=0$.

\subsection{Long - short split}

We now wish to split the quantum field into two parts: field configurations $\phi_<$ whose Green function 
is $\mathbf{D}_<$, and field configurations $\phi_>$ with corresponding Green function $\mathbf{D}_>$ such that      
\beq\label{split1}
\phi(x)=\phi_>(x) + \phi_<(x) \,, 
\eeq
and 
\beq\label{split2}
\mathbf{D}(x,x')=\mathbf{D}_>(x,x')+\mathbf{D}_<(x,x')\,.
\eeq   
We do not need to specify at this point exactly how (\ref{split2}) is achieved and the manipulations below are valid for any such split of the propagator. The fluctuations of the fields
$\phi_<$ and $\phi_>$ are by definition governed by the propagators $\mathbf{D}_<$ and $\mathbf{D}_>$ respectively. The fields $\phi_<$ and $\phi_>$ are simply dummy variables associated with path integrals weighted by $D_<$ and $D_>$ and do not need to be separately determined - see \cite{Delamotte:2007pf} for further explanation of this point. Using the following identity for Gaussian integrals
\beq
\int\limits_{-\infty}^{+\infty}  dx \, \exp\left({-\frac{x^2}{2\alpha} - V(x)}\right) = \sqrt{\frac{2\pi\beta \gamma}{\alpha}}\int  \limits_{-\infty}^{+\infty}  dydz\, \exp \left({-\frac{y^2}{2\beta} -\frac{z^2}{2\gamma} - V(y+z)} \right) \,,
\eeq
where
$x=y+z$ and $\alpha = \beta + \gamma $, we can then split the path integral as
\beq\label{Z}
Z[\boldsymbol{J}]=\int D\boldsymbol{\phi}_< D\boldsymbol{\phi}_> \exp i
\int\left[{ \frac{1}{2}\boldsymbol{\phi}_>^\dagger \mathbf{D}_>^{-1} \boldsymbol{\phi}_> + 
\frac{1}{2}\boldsymbol{\phi}^\dagger_< \mathbf{D}_<^{-1} \boldsymbol{\phi}_< - 
\overline{V}_I(\boldsymbol{\phi}_{<}+\boldsymbol{\phi}_{>})}+\boldsymbol{J}^\dagger
\left(\boldsymbol{\phi}_<+\boldsymbol{\phi}_>\right)
\right]\,d\mu\,.
\eeq
The field integrals (\ref{CTP}) and (\ref{Z}) equivalent and define the same perturbative expansions \cite{ZinnJustin:2002ru}.

We can utilize the above formal discussion for our problem by explicitly specifying the split (\ref{split2}) to reflect the separation into long wavelength and short wavelength modes. W
e can achieve this by using a window function $W(x,x')$ smoothing out short wavelength perturbations. Its 
complementary window function $\overline{W}(x,x')$, will filter out long wavelength fluctuations, satisfying  
\beq\label{sum windows}
W(x,x') + \overline{W}(x,x') ={\bf 1} \delta(x,x') \,.
\eeq
Here, long and short are defined with respect to some smoothing scale which could be both spatial and temporal. 
We normalize $\int W(x,x')\,d\mu(x') =1$ so that $\int \overline{W}(x,x') \,d\mu(x')=0$. We take the $\mathbf{D}_<$ propagator to be 
\beq\label{long} 
\mathbf{D}_<(x,x') = \int \int d\mu(x)d\mu(y) W(x,y)\mathbf{D}(y,z)W(z,x') \equiv W\mathbf{D}W\,,
\eeq
where the integrations over spacetime are implicit in the final expression. Therefore,
from (\ref{sum windows}) $\mathbf{D}_>$ is given by
\beq\label{short}
\mathbf{D}_>=  \mathbf{D}\overline{W} + \overline{W}\mathbf{D}-\overline{W}\mathbf{D}\overline{W}\,.
\eeq
For any reasonably sharp window function, $\mathbf{D}_>$ and 
$\mathbf{D}_<$ will have a negligable overlap for scales sufficiently different from the smoothing scale. 
  
Let us emphasize here that using a smooth window function $W$ does not partition the function space in 
which the field $\phi$ lives into fields with strictly short and strictly long modes. An exact step function in 
$k$-space, and only such a projector function, would be required for this. Therefore, the `integration variables' 
$\phi_<$ and $\phi_>$, integrated over in the path integral (\ref{Z}), contain \emph{all} wavelengths in their integration 
measure. However, by construction $\mathbf{D}_<$ is suppressed on short scales and hence configurations of $\phi_<$ 
with short wavelength components have suppressed contributions to the path integral due to a correspondingly 
large exponent. Similarly, the propagator (\ref{short}) ensures that long wavelength configurations do not 
contribute in the $D\phi_>$ path integral. In this sense $\phi_<$ and $\phi_>$ can meaningfully be considered 
as long and short wavelength fields respectively. Note that this formulation differs somewhat from using a direct convolution 
of $\phi$ with a window function as was originally done in \cite{Starobinsky:1986fx} 
(see also \cite{Morikawa:1989xz, Levasseur:2013ffa}), bringing this approach more in line with common 
notions of renormalization, see e.g.
\cite{Delamotte:2007pf}. More importantly, as we will see the split of the fields into long and short wavelength 
components is time dependent in inflation. Given this time dependence, the common practice of convolving 
the field with a window function does not offer itself for a clear understanding of the integration measure's 
split into long and short wavelength sectors: $D\phi\, \rightarrow \, D\phi_< D\phi_>$.

\subsection{The effective IR action}\label{eff IR act}

We now wish to integrate out the ``short wavelength'' field $\phi_>$, obtaining an effective action for 
the IR sector described by $\phi_<$. The relatively small field gradients in the IR sector
suggest that a derivative expansion in the IR fields is appropriate, and the most important
contribution to the effective action should be an effective potential term. After integrating out the high energy modes we will be left with a generating function,
\begin{equation}
Z[\boldsymbol{J}]=\int D\boldsymbol{\phi}_<\,e^{i S_<[\boldsymbol{\phi}_<]+
i\int d\mu\,\boldsymbol{J}^\dagger\boldsymbol{\phi}_{<}}\label{zeff}
\end{equation} 
with the low energy effective action defined by
\begin{equation}
e^{i S_<[\boldsymbol{\phi}_<]}=
\left\{{\exp\,i\int\frac12\boldsymbol{\phi}_{<}^\dagger \mathbf{D}_<^{-1}\boldsymbol{\phi}_{<}\,d\mu
}\right\}
\int D\boldsymbol{\phi}_>\,\exp\,i\int\left[
\frac12\boldsymbol{\phi}_{>}^\dagger \mathbf{D}_>^{-1}\boldsymbol{\phi}_{>}
-\overline V_I[\boldsymbol{\phi}_>+\boldsymbol{\phi}_<]\right]d\mu\,.
\end{equation} 
The one-loop approximation can be obtained by expanding the interaction
terms up to quadratic order,
\begin{equation}
S_I[\phi_>+\phi_<]\approx S_I[\phi_<]+\phi_{>a}{\delta S_I\over\delta\phi_{<a}}
+\frac{1}{2}\phi_{>a}{\delta^2 S_I\over\delta\phi_{<a}\delta \phi_{<b}}\phi_{>b}\,,
\end{equation}
where $S_I$ is the integral of the potential $-\overline V_I$ and repeated indices 
on the product of two functions imply an integration over spacetime.
Collecting the terms involving the high frequency fields together forms a shifted propagator $\mathbf{G}_>$,
\begin{equation}
i\mathbf{G}_{>ab}=\left(\mathbf{D}_>^{-1ab}+{\delta^2 S_I\over\delta\phi_{<a}\delta\phi_{<b}}\right)^{-1},
\label{Gshort}
\end{equation}
which depends on the slowly varying background $\phi_{<a}$.
The Gaussian integration over $\phi_>$ gives
\begin{equation}\label{Seff1}
S_<[\phi_<]\approx \int\frac12\boldsymbol{\phi}^\dagger_{<}\mathbf{D}_<^{-1}\boldsymbol{\phi}_{<}\,d\mu+ S_I[\phi_<]
-\frac{i}{2}\log\det (\mathbf{G}_>)-\frac12{\delta S_I\over\delta\phi_{<a}}\mathbf{G}_{>ab}{\delta S_I\over\delta\phi_{<b}}.
\end{equation}
The final two terms in the effective action are the leading order contributions from the short wavelength modes. 
We consider the logarithmic term below, dropping the last term as it is higher order in the non-linear coupling. Further discussion of the final term can be found in the appendix.  

The logarithm term is fairly intractable as it stands, but simplifies if we use the Keldysh representation
and expand to first order in the variable ${\phi_q}_<$, where (using the inverse metric $C_{ab}$
for convenience)
\begin{equation}
\phi_{<a}=\phi_{<}+\frac12 C_{1a}{\phi_q}_<.
\end{equation}
The first two terms in the expansion of the logarithm of (\ref{Gshort}) are,
\begin{equation}\label{log det approx}
\frac{i}{2}\log\det (G_>)\approx \frac{i}{2}\log\det (G_>[\phi])+\frac{1}{4}G_{>ab}
{\delta^3  S_I\over\delta\phi_{<a}\delta\phi_{<b}\delta\phi_{<c}}c_{1c}{\phi_q}_<.
\end{equation}
In the CTP formalism, all terms vanish when ${\phi_q}$ vanishes (see the discussion in \ref{ctpformalism}), and the first term
is therefore zero. After substituting in the interaction potential and uncondensing the notation
we have
\begin{equation}
\frac{i}{2}\log\det (G_>)\approx -\frac{1}{2}\int d\mu(x)D_{K>}(x,x)\,
{\partial^3 V_I\over\partial\phi_<{}^3}{\phi_q}_<(x).
\end{equation}
At equal time coordinate, the Keldysh propagator $D_{K}$ is equal to the shifted Feynman 
propagator $G_{>F}$,
\begin{equation}
\frac{i}{2}\log\det (G_>)\approx -\frac{1}{2}\int d\mu(x)G_{>F}(x,x)
{\partial^3 V_I\over\partial\phi_<{}^3}{\phi_q}_<(x).\label{detghf}
\end{equation}
Regularisation of the coincident limit of the Feynman propagator can be done as
described in the next section.

The low-energy effective action can be expressed in a more suggestive way by introducing the
one loop correction to the de Sitter space effective potential $\Delta V$,
\begin{equation}
\int d\mu(x) \Delta V=\frac{i}{2}\log\det G_{>F}\label{deltaV}\,.
\end{equation}
The determinant can be expanded in powers of ${\phi_q}_<$ as above, and when compared 
to (\ref{detghf}) we find
\begin{equation}
\frac{i}{2}\log\det (G_>)\approx \int d\mu(x)\left[\Delta V\left(\phi+{{\phi_q}_<\over 2}\right)
-\Delta V\left(\phi-{{\phi_q}_<\over 2}\right)\right].
\end{equation}
The quantity inside the integral is simply $\Delta V(\phi_1)-\Delta V(\phi_2)$. Recalling that $S_I$
is the integral of $V(\phi_1 )-V(\phi_2)$, we see that the logarithmic term can be included
in the effective action,
\begin{equation}
S_<[\phi_<]\approx \frac12\phi_{<a}D_<^{-1ab}\phi_{<b}+S_I[\phi_<],\label{effact}
\end{equation}
where we make the replacement $V_I\to V_I+\Delta V$ in $S_I$.

\subsection{de Sitter effective potential}\label{dS eff pot}

Calculation of the effective potential from (\ref{effact}) requires the computation of $G_F^>(x,x)$, where $G_F(x,x')$ is the 
familiar Feynman propagator. In de Sitter space computing $G_F(x,x')$ is routine, 
apart from the fact that the operator has been combined with a window function 
which suppresses the IR modes. In what follows we will express the effective potential in terms of the result for the full theory (without the window function) plus a correction term, and evaluate the latter numerically. The result for the full theory is presented first.

Consider a scalar field potential
\begin{equation}
V=\frac12m^2\phi^2+\frac14\lambda\phi^4.
\end{equation}
Equation (\ref{eqVeff}) relates the effective potential to coincident limits of the shifted Feynman propagator $G_F$. This is the free theory propagator with
mass $\tilde{m}\equiv m(\phi)$ given by
\begin{equation}
\tilde m^2(\phi)=m^2+3\lambda\phi^2.\label{effmass}
\end{equation}
The de Sitter propagator \cite{Chernikov:1968zm, Bunch:1978yq} has a closed form expression in terms of a hypergeometric function,
\begin{equation}
G_F(x,x')={\Gamma(3/2+\nu)\Gamma(3/2-\nu)H^2\over 16\pi^2}
F[3/2+\nu,3/2-\nu,2;\cos^2(H\sigma/2)],\label{desprop}
\end{equation}
where $\nu^2=9/4-\tilde m^2/H^2$. We have taken the geodesic separation $\sigma$ to be spacelike, for which $G_F$ does not have an imaginary part. 
In the limit $\sigma\to 0$, the divergent terms are removed to give the regularised coincident
limit ${\rm reg}\,G_F$,
\begin{equation}
{\rm reg}\,G_F={H^2\over 16\pi^2}\left({\tilde m^2\over H^2}-2\right)\left[\psi_{\rm poly}(3/2+\nu)
+\psi_{\rm poly}(3/2-\nu)-\ln{\mu_R^2\over H^2}\right],
\end{equation}
where ${\psi_{\rm poly}}$ is the polygamma function\footnote{Adiabatic regularisation, described for 
example in \cite{Birrell:1982ix}, removes an additional term related to the conformal anomaly.
We prefer to recover the conformal anomaly from the variation of $\mu_R$, and 
perform what might be regarded as a `minimal' subtraction in $G_F(x,x)$.}. 
Ambiguity in subtracting logarithmic divergences
results in the term containing the renormalisation scale $\mu_R$. The one loop correction to the
effective potential in de Sitter space is then
\begin{equation}\label{dvds}
\Delta V_{\rm deS}={H^2\over 32\pi^2}\int\left({\tilde m^2\over H^2}-2\right)\left[\psi_{\rm poly}(3/2+\nu)
+\psi_{\rm poly}(3/2-\nu)-\ln{\mu_R^2\over H^2}\right]\,d\tilde m^2\,,
\end{equation}
since $d\tilde{m}^2=V'''d\phi$. This indefinite integral contains an undetermined constant, but gives all the important
terms which depend on $\phi$.

In order to apply the window functions we use a spatially flat coordinate system for de Sitter, with metric
\begin{equation}
ds^2=-dt^2+a^2(t)(dx^2+dy^2+dz^2)\,.
\end{equation}
In these coordinates, the Feynman propagator has a mode expansion,
\begin{equation}
G_F(x,x')=
\begin{cases}
\sum_k u_k(t)\overline u_k(t')e^{i{\bf k}\cdot({\bf x}-{\bf x'})}&t>t'\\
\sum_k \overline u_k(t)u_k(t')e^{i{\bf k}\cdot({\bf x}-{\bf x'})}&t<t'\\
\end{cases}
\end{equation}
where $\sum_k$ is the momentum-space measure in flat space and the modes are Hankel
functions,
\begin{equation}
u_k(t)=\frac12\sqrt{\pi}H^{-1/2}a^{-3/2}H^{(1)}_{\nu}(k/aH)\,.\label{mode}
\end{equation}
We will take a window function which selects the small momentum modes,
\beq
W(t,\vc{x},t',\vc{x}')= \delta(t-t')\sum_k W_k(k,t)
e^{i\vc{k}\cdot\left(\vc{x}-\vc{x}'\right)}\label{window}
\eeq
where $W(k,t)\to0$ for large $k$.\footnote{The chosen window function smooths in space, or momentum, but not in time, hence the temporal delta function. This is the choice made in all works on stochastic inflation so far and explicitly breaks de Sitter invariance. Our formalism does not strictly require this and it would be interesting to formulate the theory using a de Sitter invariant window function which smooths time as well as space.}
The propagator with the IR modes removed is then
\begin{equation}
G^>_{F}(x,x)={\rm reg}\,G_F(x,x)-\sum_k W(k,t)^2\,u_k(t)\overline u_k(t),
\end{equation}
where the term involving the window function $W$ is finite in the coincident limit
provided that $W(t,k)$ falls off sufficiently rapidly at large $k$.
The correction to the potential for the long range modes is then
\begin{equation}
\Delta V=\Delta V_{\rm deS}-\frac12\int d\tilde m^2\sum_k W(k,t)^2\,u_k(t)\overline u_k(t).
\end{equation}
    
\begin{center}
\begin{figure}[htb]
\scalebox{0.44}{\includegraphics{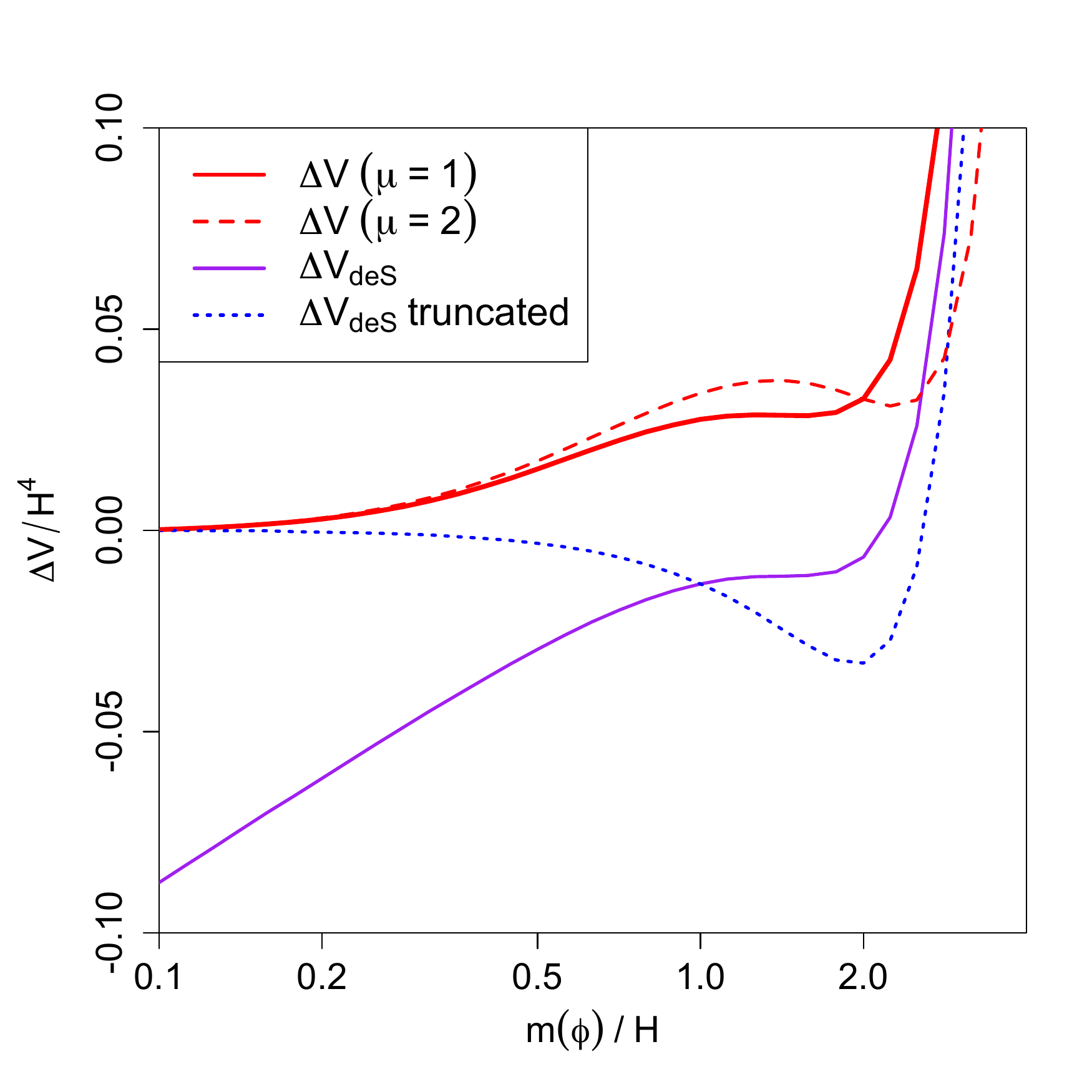}}
\scalebox{0.44}{\includegraphics{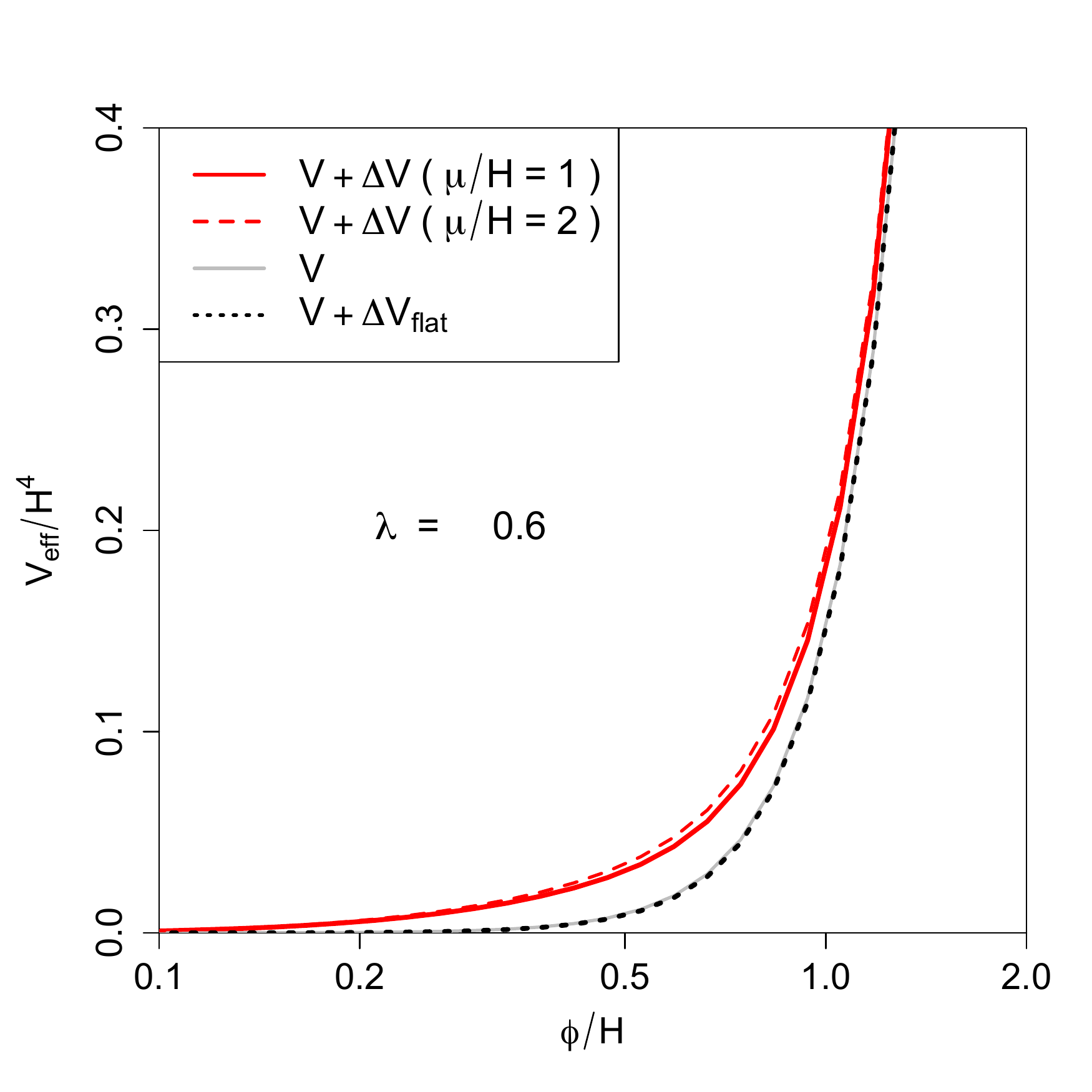}}
\caption{The left-hand plot shows the one-loop correction to the effective potential
$\Delta V$ in de Sitter space with the IR modes suppressed, as a function of the 
shifted mass. The full de Sitter result $\Delta V_{\rm deS}$ and a truncation
subtracting the leading infra-red divergence is shown for comparison. The right-hand plot 
shows the one-loop effective potential $V+\Delta V$ as a function of the field for a classical potential
$V=\lambda\phi^4/4$. The flat-space corrections are very small on this scale.}
\label{fig1}
\end{figure}
\end{center}

The effective IR potential term $\Delta V$ has been calculated numerically for a gaussian window function
\begin{equation}
W(k,t)=\exp\left(-{k^2\over a^2H^2}\right)\label{gaussian}
\end{equation}
and plotted in figure \ref{fig1}. The left hand plot is against $m(\phi)$, as defined in (\ref{effmass}),
for selected values of the renormalisation scale. The right-hand plot is specialised
to $V=\lambda\phi^4/4$. In principle, we could remove the renormalisation scale by using the 
renormalisation group to relate the running parameters in the theory to measurable parameters. 
However, since the model is not related to any particular particle model at this stage,
it is convenient to treat the renormalisation scale as an extra parameter.

The left-hand plot in figure \ref{fig1} shows the effect of suppressing the IR modes in the short wavelength propagator 
contributing to the effective potential. The full de Sitter result has a logarithmic divergence in the limit 
$\tilde m\to 0$,
\begin{equation}
\Delta V_{\rm deS}\sim {3H^4\over 16\pi^2}\ln{\tilde m^2\over H^2}-{\tilde m^2H^2\over 16\pi^2}
\left\{\ln{H^2\over\mu_R^2}-2\gamma+\frac{10}{3}\right\}+\dots,\label{dvmsmall}
\end{equation}
where $\gamma=0.577216\dots$ is Euler's constant.
Suppressing the IR modes has done the job of removing the leading logarithmic $H^4\ln m^2$ term.
At the opposite extreme of mass, the plot rapidly approaches the flat-space result as 
$m$ is increased beyond $m=H$,
\begin{equation}
\Delta V_{\rm flat}={\tilde m^4\over 64\pi^2}\left\{\ln{\tilde m^2\over\mu_R^2}-\frac12\right\},
\end{equation}
The logarithm in the flat space result becomes large for small mass, leading to
a breakdown in perturbation theory. This can be corrected by using the renormalisation
group corrected potential.

The result of truncating the full de Sitter one loop potential by simply dropping the 
logarithmic term is plotted in $\Delta V_{\rm deS}$ truncated. There is a significant
qualitative difference between this and $\Delta V$, which shows that suppressing the IR modes, as we have done, 
does more than just remove the large logarithmic terms in the 
potential. The leading order behaviour of the truncated theory is given by the second
term in (\ref{dvmsmall}), which is negative for $\mu_R\sim H$. Suppressing the IR
modes gives a positive $\tilde m^2$ term for $\mu_R\sim H$ which dominates the one-loop 
potential in the right-hand plot as $\phi\to 0$. The value of the effective mass at
$\phi=0$ can be varied by changing the renormalisation scale.


\subsection{The IR dynamics}

In order to complete our construction of the IR dynamics we also need a convenient representation for the operator 
inverse of $\mathbf{D}_<$ which appears in the effective action (\ref{Seff1}). We can exploit the fact that 
de Sitter correlations decay at most with a mild power law at large distances \cite{Garbrecht:2014dca}, 
while $\mathbf{D}_>$ is by construction small above the smoothing scale. Hence, when acting on long 
wavelength fields, we can formally write
\bea\label{expansion}
\mathbf{D}_<^{-1}=\frac{1}{\mathbf{D}-\mathbf{D}_>}
&=&\mathbf{D}^{-1} + \mathbf{D}^{-1}\mathbf{D}_>\mathbf{D}^{-1}+\ldots\nonumber\\
&=&\mathbf{D}^{-1}  -  \mathbf{D}^{-1}\overline{W}\mathbf{D}\overline{W}\mathbf{D}^{-1}
+\mathbf{D}^{-1}\overline{W}+ \overline{W}\mathbf{D}^{-1}+\ldots
\eea   
where in the second line we used (\ref{short}) and the operator series on the r.h.s is understood to 
act on functions with only long wavelength support. Thus we find that the dominant contribution to the long 
wavelength action, expressed in the 
Keldysh basis, is 
\beq\label{quadratic LW}
\boldsymbol{\phi}^\dagger\mathbf{D}_<^{-1} \boldsymbol{\phi} \simeq
\left(\begin{matrix} \phi_c,&{\phi_q}\end{matrix}\right)
\left(\begin{matrix}0& \mathfrak{D} \\ \mathfrak{D} & 0	
\end{matrix}\right)\left(\begin{matrix}\phi_c\\ {\phi_q}\end{matrix}\right) 
- \left(\begin{matrix} \phi_c,&{\phi_q}\end{matrix}\right)\left(
\begin{matrix}
0& \mathfrak{D} \overline{W}D_A\overline{W}\mathfrak{D} \\
\mathfrak{D}\overline{W}D_R\overline{W}\mathfrak{D}& -i\mathfrak{D}\overline{W}D_K\overline{W}\mathfrak{D}
\end{matrix}
\right)\left(\begin{matrix}\phi_c\\ {\phi_q}\end{matrix}\right) \,,
\eeq  
where $\mathfrak{D}=\nabla^2-m^2$. We have removed the subscript $<$ to simplify notation, and 
spacetime integrations are implicit in the elements of the second matrix. Any terms where 
$\overline{W}$ is directly convolved with $\phi$ only provide higher order 
spatial gradient terms which are subdominant and were dropped from (\ref{quadratic LW}). Noting that $D_A$ and $D_R$ are suppressed on long wavelengths compared to the Keldysh propagator $D_K$, the dominant contribution to the second term of (\ref{quadratic LW}) comes from the ${\phi_q}^2$ term. A feature of the Keldysh formalism is that such terms
are equivalent to adding a stochastic force 
$\xi(x)$ \cite{Altland:2006si} to the equation of motion with correlator  
\beq
\langle\xi(x)\xi(x')\rangle \equiv\mc{N}({x},{x}')= 
\int\mathfrak{D}_{x}\mathfrak{D}_{x'}\overline{W}(x,y)D_K(y,z)\overline{W}(z,x')d\mu(y)d\mu(z).\label{noisedef}
\eeq
This corresponds to the well known Starobinsky noise term \cite{Starobinsky:1986fx}. 

Rescaling $\psi=a^3{\phi_q}$, the exponent in the path integral of the IR theory is written as 
\bea\label{MSRJD-1}
iS[\phi,{\phi_q}]=&&i\int dt d^3{\vc{x}} \,\Bigg[\frac{1}{2}
\left(\begin{matrix} \phi,&\psi\end{matrix}\right)\left(\begin{matrix}
	0& (\nabla^{2\dagger}-{m}^2) \\ (\nabla^2-{m}^2) &
	 -i\int_{x'}\mc{N}(x,x')	
\end{matrix}\right)
\left(\begin{matrix}\phi\\ \psi
\end{matrix}\right)-\frac{\partial V_{I \rm eff}}{\partial\phi}\psi\nonumber\\ &&
+2\!\sum\limits_{m=1}^\infty\!\frac{V_{I\rm eff}^{{(2m+1)}}}{\left(2m+1\right)!}
\left(\frac{\psi}{2}\right)^{2m+1}\!\left( \frac{H}{a}\right)^{6m}\Bigg]
\eea
where 
\bea\label{nabla sq 1}
-\nabla^2 &=&\partial_t^2+3H\partial_t-H^2a^{-2}\partial^2_{\bf x}\\
\label{nabla sq 2}-\nabla^{2\dagger} &=&\partial_t^2-3\partial_tH-H^2a^{-2}\partial^2_{\bf x}\,.
\eea
Note that the anti-damping in (\ref{nabla sq 2}) results from absorbing the $a^3$ factor in $\psi$ which makes the operator non-self-adjoint. It is clear that powers of $\psi$ higher than $2$ are highly suppressed by inverse powers of $a$, inducing 
semi-classical dynamics in the IR. The only remnant of quantum fluctuations is the stochastic $\psi^2$ term. 
Note that this argument holds for the fully interacting theory, relying on the \emph{growth of the volume} 
of the system and not on the behaviour of the linear mode functions on long wavelengths \cite{Polarski:1995jg}, 
although the latter is related to the former. This justifies the approximation we made earlier in (\ref{log det approx}) 
as the neglected terms would end up being suppressed. We have thus arrived at the following action on long wavelengths 
\beq\label{MSRJD-2}
iS[\phi,\psi]\simeq -i\int dt\, d^3\vc{x}
\left[\psi\left(-\nabla^2\phi+{m}^2\phi+\frac{\partial V_{I \rm eff}}{\partial\phi}\right)\right]  
-\frac{1}{2}\int d^3\vc{x} d^3\vc{x}'\,dtdt' \,\psi(x)\mc{N}(x,x')\psi(x'),
\eeq
which describes stochastic Langevin dynamics for $\phi$
\beq\label{langevin}
-\nabla^2\phi+{m}^2\phi+\frac{\partial V_{I\rm eff}}{\partial\phi}=\xi(t,\vc{x})
\eeq
with the $\xi$ correlator given by (\ref{noisedef}). 

The noise kernel $\mc{N}(x,x')$ depends of course on the window function but any physical 
results should be independent of this choice. The original formulation of stochastic inflation by Starobinsky, 
which neglected the field acceleration $\ddot{\phi}$, used a sharp step function in $k$ space to define the 
long wavelength system. Here we use
the gaussian window function,
\begin{equation}
W(k,t)=\exp\left(-{k^2\over a^2H^2}\right).\label{gaussian}
\end{equation}
In momentum space, the noise kernel (\ref{noisedef}) becomes
\beq
{\cal N}({x},{x}')= 
\sum_ke^{i{\bf k}\cdot({\bf x}-{\bf x}')}
\mathfrak{D}_{t}\mathfrak{D}_{t'}\overline{W}(k,t)u_k(t)\bar u_k(t')\overline{W}(k,t').
\eeq
where
\begin{equation}
\mathfrak{D}=-\partial_t^2-3H\partial_t-a^{-2}k^2-m^2.
\end{equation}
The window functions pick out the small $k$ limit, where we can approximate the modes (\ref{mode}) by,
\begin{equation}
u_k(t)\bar u_k(t')\approx
\frac{H^2}{2k^3} \left(\frac{k}{a(t)H}\right)^{\frac{m^2}{3H^2}}\left(\frac{k}{a(t')H}\right)^{\frac{m^2}{3H^2}}.
\end{equation}
For small mass $m\ll H$, the integration results in
\beq
{\cal N}({x},{x}')= 
{3H^6\over 2\pi^2}{\rm sech}^4 \,H(t-t')\,
{}_1F_1(-5/2,3/2;\kappa^2|{\bf x}-{\bf x}'|^2)e^{-\kappa^2|{\bf x}-{\bf x}'|^2},
\eeq
where ${}_1F_1$ is a confluent hypergeometric function and
\begin{equation}
\kappa=\frac12H\,{\rm sech} \,H(t-t')\,e^{H(t+t')/2}.
\end{equation}

We can get an understanding about the IR behaviour in various limits. For sub-horizon scales
$|{\bf x}-{\bf x}'|<1/(aH)$, the noise resembles white noise with a smoothed delta-function
kernel,
\begin{equation}
{\cal N}({x},{x}')\sim {9H^5\over 4\pi^2}\delta(t-t')\,.\label{noiseapp}
\end{equation}
For super-horizon separations the noise correlations are exponentially suppressed. This behaviour is repeated for other types of window functions. Hence, on large scales, $k<aH$, evolution becomes quasi-local in space and different regions of physical size $\Delta r \sim 1/H$ are acted upon by spatially uncorrelated noise $\xi$. We can thus approximate
\beq
{\cal N}(x,x') \simeq N({\bf x},{\bf x}',t)\,\delta(t-t'),
\eeq
where the spatial kernel is
\beq
N({\bf x},{\bf x}',t) \simeq  {9H^5\over 4\pi^2}\theta\left(1-aH\left|\vc{x}-\vc{x}'\right|\right)\label{defNker}.
\eeq

Having found the noise statistics, we shall now examine the long-range dynamics for the free field theory.
Since only $k<aH$ modes are included in $\mathbf{D}_<$, the propagator functions become approximately 
local in space \cite{Woodard:2005cw}. If we let ${\cal G}^R(t,t')$ be the Green function for the stochastic system (\ref{langevin})
dropping the gradient terms, then
\beq\label{Dk<}
D_K^<(x,x') = \frac{9 H^5}{4\pi^2}\int\limits_{0}^{+\infty} d\tau \, \mc{G}^R(t,\tau)\mc{G}^A(\tau,t')\theta\left(1-a(\tau)H\left|\vc{x}-\vc{x}'\right|\right) + \mc{F}_0(x,x')
\eeq
where ${\cal F}_0$ is a function satisfying the equation of motion, resulting from the arbitrariness of inverting $D_<^{-1}$. 
Its choice reflects a choice of initial state. For $\Delta\vc{x}=0$, (\ref{Dk<}) gives the coincident correlation function as
\beq\label{corr2}
\langle\phi(t,\vc{x})\phi(t',\vc{x})\rangle = D_K^<(t,\vc{x},t',\vc{x})=
\frac{3 H^4}{8\pi^2m^2}\left(e^{-\frac{m^2}{3H}|t-t'|} -\frac{m^2}{9H^2}e^{-3H|t-t'|} \right)\,,
\eeq
where we chose ${\cal F}_0$ such that the correlation function assumes a unique form depending only on $|t-t'|$. 
For non-zero comoving separation $\Delta\vc{x}$ we have
\beq\label{corr3}
D_K^<(x,x') = \frac{9 H^5}{4\pi^2}\int_{0}^{t_\star} d\tau \, \mc{G}^R(t,\tau)\mc{G}^A(\tau,t')+\mc{F}_0(t,t',|\Delta\vc{x}|)
\eeq   
where $t_\star = -H^{-1}\ln \left(H\left|\Delta \vc{x}\right|\right)$, and we assumed that $H\left|\Delta \vc{x}\right|<1$, i.e.~the 
points are initially within the horizon. Regardless of the specific choice of $\mathcal{F}_0$, relation (\ref{corr3}) reproduces 
via stochastic dynamics the asymptotic result for free quantum fields in the Bunch-Davies de Sitter vacuum, 
\beq
\langle\phi(t,\vc{x})\phi(t',\vc{x}')\rangle \sim \frac{3H^4}{8\pi^2 m^2}\left(a(t)a(t')H^2\left|\Delta \vc{x}\right|^2\right)^{-\frac{m^2}{3H^2}} \,.
\eeq

\section{The Kramers Equation for $\phi$}

The benefits of using the stochastic dynamics become apparent when we consider the interacting theory.
It is possible to find probability distributions for $\phi$ at a spatial point ${\bf x}$ and use these to 
calculate non-perturbative expectation values of functions of $\phi$ in the stochastic theory.
By the previous construction, these will be excellent approximations to the expectation
values of the full quantum theory. As a further step, which is less well known,
one can construct joint probability distributions of $\phi$ and $\dot\phi$ in the 
stochastic theory in the form of a Wigner function which satisfies a Kramers equation. Kramers equations were first applied for the study of cosmological IR fluctuations in \cite{Riotto:2011sf}.     

As a starting point, recall the definition of the Wigner function for a coordinate
$x$ and momentum $p$ with density matrix $\hat\rho(t)$,
\begin{equation}
W(x,p,t)={1\over 2\pi \hbar}\int_{-\infty}^\infty\,dq
\left\langle x+\frac{q}{2}|\hat\rho(t)|x-\frac{q}{2}\right\rangle e^{-ipq/\hbar}.
\end{equation}
For the IR dynamical field we introduce the Wigner functional using a path integral over 
spatial fields $\psi$,
\begin{equation}
{\cal W}[\phi,v,t]={1\over 2\pi}\int_{-\infty}^\infty D\psi\,\rho[\phi,\psi,t]
e^{-i\int d^3{\vc{x}} \,\psi\,v}.\label{wignertransform}
\end{equation}
We have found that the the leading-order IR dynamics is 
given by a path integral over the long-range spacetime fields $\phi'$ and $\psi'$, where
the prime will be used now to clearly distinguish spacetime fields from
purely spatial fields. 
We will keep careful track of the field values held fixed at time $t$, and add these as superscript indices
on the integral sign. The density matrix is given by a path integral
\begin{equation}
\rho[\phi,\psi,t]=\int^{\phi,\psi}D\phi' D\psi' \,K[\psi']\,e^{-i\int_{t_i}^t d^4x
\left[\psi'(\ddot\phi'+3H\dot\phi'-a^{-2}\boldsymbol{\nabla}^2\phi'+dV/d\phi')\right]
+i\int d^3\vc{x}\psi\dot\phi},
\end{equation}
where spatial coordinates have been suppresed and the fields $\phi'(t)=\phi$ and 
$\psi'(t)=\psi$ are held fixed at time $t$.
The noise kernel
\beq
K[\psi]=\exp\left(-\frac12\int d^4xd^4x' \psi(x)\mathcal{N}(x,x')\psi(x')\right),
\eeq
with $\mathcal{N}(x,x')$ given by (\ref{noiseapp}). Normalisation constants are dropped
since they play no further role.
The boundary conditions on the fields are also set by some predetermined initial
conditions at time $t_i$ as well as the fixed field values at time $t$. Because the Lagrangian
is second order in derivatives, the boundary conditions require the addition of the boundary term 
$\psi\dot\phi$ in the exponent to cancel boundary terms from the variation of the Lagrangian.

The Wigner function satisfies a Fokker-Planck equation, which we will now obtain.
First, introduce an auxiliary spacetime field $v$ to replace $\dot \phi$ and a Lagrange
multiplier field $z$,
\begin{equation}
\rho[\phi,\psi,t]=\int^{\phi,\psi}D\phi' D\psi' Dv' D{z'} \,K[\psi']\,e^{-i\int_{t_i}^t d^4x
\left[{z}(\dot\phi'-v')-\psi'(\dot v'+3Hv'+\delta U/\delta\phi')
\right]
+i\int d^3\vc{x}\psi(t)v},
\end{equation}
where we introduced the total potential energy
\beq
U=\int d^3\vc{x} \left(V_{\rm eff} +\frac{1}{2}\frac{(\boldsymbol{\nabla}\phi)^2}{a^2}\right)\,.
\eeq
Note that eq. (\ref{wignertransform}) for the Wigner functional integrates out the field $\psi$, so this is no 
longer fixed. However, the auxiliary field $v$ does not get integrated, hence
\begin{equation}
{\cal W}[\phi,v,t]=\int^{\phi,v}D\phi' D\psi' Dv' D{z'} \,K[\psi']\,e^{i\int_{t_i}^t d^4x
\left[{z'}(\dot\phi'-v')-\psi'(\dot v'+3Hv'+\delta U/\delta\phi')
\right]}\,
\end{equation}
The path integral can then be put into canonical form,
\begin{equation}
{\cal W}[\phi,v,t]=\int^{\phi,v}D\phi' D\psi' Dv' Dz' \,e^{i\int_{t_i}^t d^4x
\left[{z'}\dot\phi'-\psi' \dot v'\right]-i\int_{t_i}^t dt\,H_{\rm stoc}},
\end{equation}
which has momenta $({z},-\psi)$ and a ``pseudo-Hamiltonian''
\beq
H_{\rm stoc}=\int d^3\vc{x}d^3\vc{x}'\,\mathcal{H}(\psi,{z},v,\phi),
\eeq
where
\beq
{\cal H}=
\left[\psi\left(3Hv+\frac{\delta U}{\delta \phi}\right)+zv\right]\delta({\bf x}-{\bf x}')-
\frac{i}2\psi(t,\vc{x})N(\vc{x},\vc{x}')\psi(t,\vc{x}')\,.
\eeq

By setting $\psi=i\delta/\delta v$ and ${z}=-i\delta/\delta\phi$ and choosing an operator ordering in the 
pseudo-Hamiltonian, the Wigner function $\mc{W}$ will satisfy a corresponding ``Schr\"{o}dinger'' equation
$i\partial_t{\cal W}={\cal H}{\cal W}$, or equivalently
\bea\label{FP1}
\partial_t\mathcal{W}&=&
\frac12\int d^3\vc{x}d^3\vc{x}'N({\bf x},{\bf x}')\frac{\delta^2\mc{W}}{\delta v(\vc{x})\delta v(\vc{x}')}\nonumber\\
&+&
\int d^3{\vc{x}}\left[3H\frac{\delta}{\delta v(\vc{x})}v(\vc{x})+\frac{\delta U}{\delta\phi(\vc{x})}\frac{\delta}{\delta v(\vc{x})}-
v(\vc{x})\frac{\delta}{\delta\phi(\vc{x})}\right]\mc{W}\,.
\eea
This is a functional equation, structurally similar to a Kramers equation \cite{Risken}, for the field $\phi$ and its velocity $v$. 
It is clear that for any fixed comoving separation $\Delta\vc{x}$ the diffusion term only acts for a finite time with 
the classical drift terms dominating the later evolution. Hence when fixed comoving points are dragged outside 
the horizon they no longer experience stochastic diffusion relative to each other, allowing for long wavelength 
correlations such as the ones observed in the CMB to be preserved. Continuous action of the noise takes 
place only for $\Delta\vc{x}=0$, or for fixed physical sub-horizon separations.

Expectation values of functions of $\phi(\vc{x})$ and $v(\vc{x})$ can be computed using $\mc{W}$.
The coincident field variance  
\beq
\langle\phi^2(\vc{x})\rangle_{\rm stoc}\equiv \int \mc{D}v \mc{D}\phi \,\phi^2(\vc{x}) \mc{W}
\eeq
is of particular interest. Using (\ref{FP1}) it can be shown to satisfy  
\beq\label{stochexp}
\left(\partial_t^2+3H\partial_t
\right)\left\langle\phi^2\right\rangle_{\rm stoc}
=2\left\langle v^2 -\frac{1}{a^2}\left(\boldsymbol{\nabla}\phi\right)^2 \right\rangle_{\rm stoc}- 
2\left\langle\phi\frac{dV_{\rm eff}}{d\phi}\right\rangle_{\rm stoc} \,,
\eeq
where we used $\boldsymbol{\nabla}\langle\ldots(\vc{x})\rangle_{\rm{stoch}}=0$ for any coincident stochastic expectation value computed with a probability distribution functional of $\phi(\vc{x})$.
We thus see that the stochastic IR theory formally reproduces the operator equation
\beq\label{qftexp}
-\nabla^2\langle\hat{\phi}^2\rangle = 2\langle -\nabla_\mu\hat\phi\nabla^\mu\hat\phi\rangle 
- 2\left\langle\hat{\phi}\frac{d{V}}{d\hat\phi}\right\rangle \,,
\eeq  
resulting from the operator equations of motion for $\hat{\phi}$, for translationally invariant states with $\boldsymbol{\nabla}\langle\hat{\phi}^2\rangle =0$. Note that in the stochastic version $V$ is replaced by the long wavelength effective potential $V_{\rm eff}$ and operator averages are replaced by probabilistic expectation values weighted by $\mc{W}[\phi(\vc{x})]$. To further demonstrate the correspondence between the stochastic and the quantum theory, we need to compare the first terms on the rhs of (\ref{stochexp}) and (\ref{qftexp}). We do this below, after we derive a simplification to (\ref{FP1}).    

The full solution to equation (\ref{FP1}) contains information for expectation values of field products at separate spacetime points. Leaving a detailed study for future work, we obtain here a simpler expression that allows for the computation of \emph{coincident} expectation values. We can use a derivative expansion of the non-local Wigner functional ${\cal W}$, keeping just the leading term
\beq\label{leading}
\ln{\cal W}\approx \omega \int d^3{\bf x}\,\ln W(\phi,v,t),
\eeq
where $\omega$ is a constant factor, roughly the inverse of the volume over which gradients have been smoothed out $\omega \sim H^{-3}$. This approximation should be adequate for computing coincident expectation values since in the IR theory, and for points closer than $\Delta r \sim H^{-1}$, gradient terms are negligible by construction since there are no short wavelength modes. We then have
\beq
{1\over {\cal W}}{\delta {\cal W}\over \delta \phi}={\omega\over W}{\partial W\over\partial\phi}\,.
\eeq
and
\beq\label{d2Wdvdv}
{1\over{\cal W}}{\delta^2 {\cal W}\over \delta v({\bf x})\delta v({\bf x}')}={\omega\over W^2}
\left(\omega{\partial W\over\partial v({\bf x})}{\partial W\over\partial v({\bf x}')}
-{\partial W\over\partial v({\bf x})}{\partial W\over\partial v({\bf x}')}\delta({\bf x}-{\bf x}')\right)
+{\omega\over W}{\partial^2 W\over\partial v({\bf x})^2}\delta({\bf x}-{\bf x}').
\eeq
The fact that the spatial gradients of the IR field are negligible on subhorizon scales results in a spatial localisation
of the noise kernel in (\ref{defNker}), allowing for a simplification of the ``diffusion'' term. We can
pick the constant $\omega$ such that the first two terms in (\ref{d2Wdvdv}) cancel to leading order in spatial gradients and then 
\beq
{1\over{\cal W}}
\int d^3\vc{x}d^3\vc{x}'N({\bf x},{\bf x}')\frac{\delta^2\mc{W}}{\delta v(\vc{x})\delta v(\vc{x}')}
\approx {9H^5\over 4\pi^2}\int d^3{\bf x}{\partial ^2W\over \partial v^2}
\eeq
The equation for the Wigner functional thus reduces to an ordinary Kramers equation for $W$,
\beq\label{FP2}
\partial_t{W}=
\left[\frac{9H^5}{8\pi^2}\frac{\partial^2}{\partial v^2}+
3H\frac{\partial}{\partial v}v+\frac{dV_{\rm eff}}{d \phi}
\frac{\partial}{\partial v}-
v
\frac{\partial}{\partial\phi}
\right]{W}\,.
\eeq
Interestingly, (\ref{FP2}) has been found previously by starting from general properties of a one-dimensional stochastic field in de Sitter space \cite{Graziani:1988,Buryak:1995tx}. Kramers equations were also used to study IR fluctuations in inflation in \cite{Riotto:2011sf}.  
We have arrived at this result by a series of steps starting from the full quantum field theory in de Sitter 
space. 

Let us now remark on the evolution of $\langle \phi^2\rangle$ as described by QFT and the stochastic IR theory. In QFT, in the de Sitter invariant Bunch-Davies vacuum, we have the renormalized value $\langle \nabla_\mu\hat\phi\nabla_\nu\hat\phi\rangle = -g_{\mu\nu}{3H^4}/{32\pi^2}$ \cite{Vilenkin:1982wt} which implies     
\beq\label{qftexp2}
\left(\partial_t^2+3H\partial_t\right)\left\langle\hat{\phi}^2\right\rangle=
\frac{3 H^4}{4\pi^2}- 2\left\langle\hat{\phi}\frac{dV}{d\hat{\phi}}\right\rangle\,.
\eeq
The corresponding result is automatically obtained in our stochastic formalism. One notes that on a timescale $\Delta t \sim H^{-1}$ the stochastic velocity $v$ reaches an equilibrium distribution 
\beq\label{MB-dist1}
W(\phi,v)\propto e^{-\frac{\Omega}{2}v^2},
\eeq 
where $\Omega=8\pi^2/3H^4$, implying that
\beq
\left\langle v^2\right\rangle_{\rm eq}=\frac{3 H^4}{8\pi^2}\,.
\eeq
Since gradient terms are absent in the stochastic formalism for subhorizon separations, we have 
\beq\label{stochexp2}
\left(\partial_t^2+3H\partial_t\right)\left\langle\phi^2\right\rangle_{\rm stoc}=
\frac{3 H^4}{4\pi^2}- 2\left\langle\phi\frac{dV_{\rm eff}}{d\phi}\right\rangle_{\rm stoc} \,,
\eeq
in correspondence to the QFT result. We see that including the stochastic velocity $v$ naturally recovers the correct growth of the field variance $\langle\hat{\phi}^2\rangle$ \cite{Vilenkin:1982wt, Linde:1982uu, Starobinsky:1982ee, Vilenkin:1983xp} within the stochastic formalism, with $\langle v^2\rangle_{\rm eq}$ replacing the QFT $\langle(\nabla\hat{\phi})^2\rangle$ term in the Bunch-Davies vacuum. It is worth noting that the relation $\langle \nabla_\mu \phi \nabla_\nu \phi \rangle = \frac{1}{4}g_{\mu\nu}\langle(\nabla\phi)^2\rangle$ cannot be imposed for classical stochastic fields as it would imply $\langle\dot\phi^2\rangle = - \frac{1}{3}\langle(\partial_i\phi)^2\rangle$ which is impossible for expectation values of positive quantities with respect to a probability distribution. One is then led to conclude that either a) the stochastic formalism could be amended to impose this symmetry, or b) an effective decoherence occurs which necessarily breaks de Sitter symmetry \cite{Rigopoulos:2013exa, Clifton:2014fja, Markkanen:2016jhg}.     

Equation (\ref{FP2}) admits a unique normalizable time-independent equilibrium solution 
\beq\label{MB-dist}
W(\phi,v)=\mc{A}e^{-\Omega\left(\frac{1}{2}v^2+V_{\rm eff}\right)},
\eeq  
where ${\cal A}$ is a normalisation constant. At equilibrium 
\beq\label{equipartition}
\left\langle v^2\right\rangle_{\rm eq}=
\left\langle\phi\frac{dV_{\rm eff}}{d\phi}\right\rangle_{\rm eq} = \frac{3 H^4}{8\pi^2}\,,
\eeq 
as it should be from (\ref{stochexp2}).\footnote{Note that this implies equipartition $\langle V\rangle = \frac{1}{2}\langle v^2\rangle_{\rm eq}$ 
for a free field, which is consistent with a thermal nature for the state.} Full equilibrium is established over much longer timescales that depend on the parameters of the potential. For example,
the time to reach the equilibrium distribution $\Delta t_{\rm equil} \sim H/m^2$ for $V= \frac12m^2\phi^2$ or $\Delta t_{\rm equil} \sim 1/\sqrt{\lambda}H$ for $V=\frac14 \lambda \phi^4$. It would be interesting to check these timescales against a one-loop QFT calculation.

\section{The stress-energy tensor}

As an example of the stochastic formalism and the use of the Wigner function, we shall consider the expectation value of the 
stress-energy tensor for a light scalar field. We therefore introduce
\begin{equation}
\langle T_{\mu\nu}\rangle=\left\langle\nabla_\mu\phi\nabla_\nu\phi-
g_{\mu\nu}\left[\frac12(\nabla\phi)^2+V\right]\right\rangle\,.
\end{equation}
It immediately becomes apparent that the full solution to (\ref{FP1}), also containing information about the spatial variation of 
$\phi$ is needed to reliably calculate the expectation values of spatial derivative terms using the stochastic formalism. 
However it is possible to reach some conclusions using the reduced Wigner function (\ref{MB-dist}) assuming 
that the system finds itself in a de Sitter invariant state. We will also comment on the ways in which the field can be thought of 
developing a `dynamical mass' \cite{Starobinsky:1994bd, Garbrecht:2011gu, Beneke:2012kn} from the interactions. 

When discussing expectation values at coincident spacetime points, de Sitter invariance corresponds to using the 
equilibrium distribution (\ref{MB-dist}), since de Sitter invariant expectation values are time independent. We can then write for the expectation value of $T_{\mu\nu}$ 
\begin{equation}\label{dST}
\langle T_{\mu\nu}\rangle=\frac14 g_{\mu\nu}\langle T_\rho{}^\rho\rangle\,.
\end{equation}
We can also express the expectation value as a functional derivative
\begin{equation}
\langle T_\rho{}^\rho\rangle=
g_{\rho\sigma}{2i\over Z}{\delta Z\over\delta g_{\rho\sigma}},
\end{equation}
where we previously reduced $Z$ to the generating functional for the IR theory in (\ref{zeff}). We have to
take a little care because the regularisation has introduced a scale dependence,
so that there are long-range and short-range contributions to the expectation value,
\begin{equation}
\langle T_\rho{}^\rho\rangle=\langle T_\rho{}^\rho\rangle_{\rm eq}
+\langle T_\rho{}^\rho\rangle_{\rm anom}.
\end{equation}
The first term above is to be computed using the equilibrium probability distribution or Wigner functional while the 
second term is an anomalous term related to the variation of the renormalization scale. Variation of the metric in 
the effective action gives\footnote{In the CTP formalism, this involves
introducing two metrics $g_{1\mu\nu}$ and $g_{2\mu\nu}$ associated with the fields $\phi_1$ and
$\phi_2$.}
\begin{equation}
\langle T_\rho{}^\rho\rangle_{\rm eq}=\langle-(\nabla\phi_<)^2-4V(\phi_<)\rangle,
\end{equation}
and these IR terms can be evaluated using the stochastic theory. As we discussed above, the spatial gradients are absent in the stochastic 
formalism when computing coincident expectation values and, numerically, $\langle\nabla_\mu\hat{\phi}\nabla^\mu\hat{\phi}\rangle=-\langle v^2\rangle_{\rm eq}$. We are thus left with just
\begin{equation}
\langle T_\rho{}^\rho\rangle_{\rm eq}=\langle v^2-4V\rangle_{\rm eq}=
\langle\phi V'-4V\rangle_{\rm eq},
\end{equation}
where we noted that the equilibrium distribution (\ref{MB-dist}) implies that
\beq
\langle v^2\rangle_{\rm eq}=\langle\phi V'\rangle_{\rm eq}\,.
\eeq 

The anomalous trace is given by variation of the renormalisation scale
\begin{equation}
\langle T_\rho{}^\rho\rangle_{\rm anom}=\mu_R{d\over d\mu_R}\Delta V_{\rm deS},
\end{equation}
for $\Delta V_{\rm deS}$ given in (\ref{dvds}). The full result for
the anomalous trace of a free scalar field of mass $\tilde m$ is \cite{Birrell:1982ix},
\begin{equation}
\langle T_\rho{}^\rho\rangle_{\rm anom}={1\over 16\pi^2}
\left(\frac1{15}H^4-\frac12(\tilde m^2-2H^2)^2\right).
\end{equation}
If we use a shifted mass $\tilde m$ depending on $\phi$, then we take the
stochastic average of the anomalous trace.

\begin{figure}[t]
\begin{center}
	\scalebox{0.45}{\includegraphics{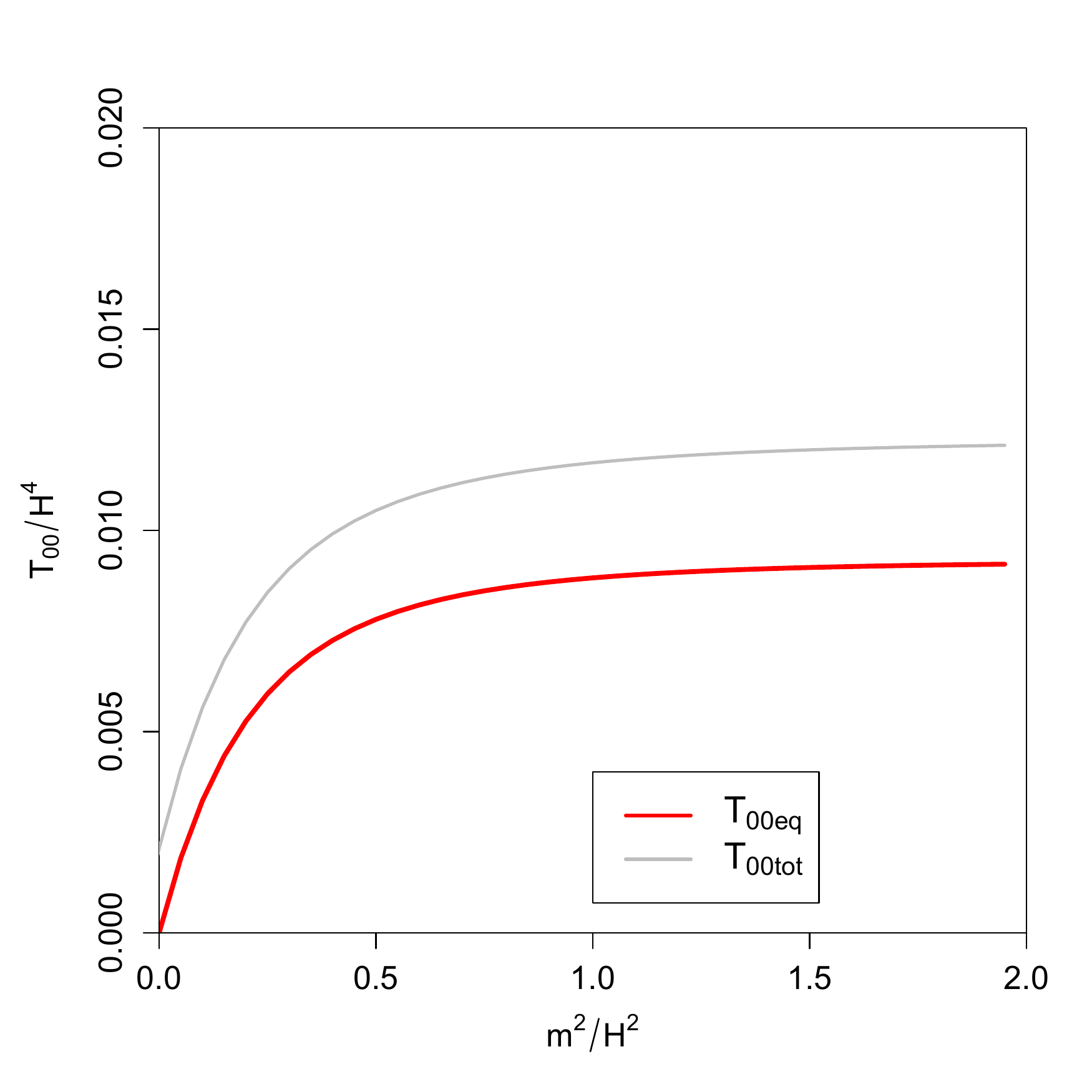}}
\caption{The plot shows the value of $\langle T_{00}\rangle$ for a light scalar field of mass $m$
and $\lambda=0.6$ in de Sitter space with expansion rate $H$. The total result is the sum of 
the stochastic theory $\langle T_{00}\rangle_{\rm eq}$ and the anomalous trace contribution.
}
\label{fig2}
\end{center} 
\end{figure}

For a specific computation, consider an effective long range potential
\begin{equation}
V_{\rm eff}=\frac12 m^2\phi^2+\frac14\lambda\phi^4
\end{equation}
with a small effective mass $m\ll H$. In this case,
\begin{equation}
\langle T_{\mu\nu}\rangle_{\rm eq}=-\frac14g_{\mu\nu}m^2\langle\phi^2\rangle_{\rm eq}
\end{equation}
where from (\ref{MB-dist})
\begin{equation}
\langle\phi^2\rangle_{\rm eq}={1\over Z}\int_{-\infty}^\infty d\phi\,\phi^2e^{-\Omega V}
\end{equation}
and $Z$ is a normalisation constant.
The values of $\langle T_{00}\rangle_{\rm eq}$ and $\langle T_{00}\rangle$ have 
been plotted in figure \ref{fig2}, for the case $\lambda=0.6$ and $\tilde m^2=3\lambda\phi^2$.
The effective mass can be varied by changing the renormalisation scale.

Approximate results can be obtained by introducing
\begin{equation}
I(g)={1\over \sqrt{2\pi}}\int_{-\infty}^\infty dx\,e^{-x^2/2-gx^4},
\end{equation}
then
\begin{equation}
\langle\phi^2\rangle_{\rm eq}={3H^4\over 8\pi^2m^2}+{3H^4\over 2\pi^2m^2}{d\ln I\over d\ln g},
\end{equation}
where
\begin{equation}
g={3\lambda H^4\over 32 m^4}.
\end{equation}
The leading terms in $\langle\phi^2\rangle$ are
\begin{equation}
\langle \phi^2\rangle_{\rm eq}\sim
\begin{cases}
\displaystyle{3H^4\over 8\pi^2m^2}&3\lambda\ll32\pi^2 m^4/H^4\\[2ex]
\displaystyle{\sqrt{3}\Gamma(3/4)^2\over 2\pi^2}{H^2\over \sqrt{\lambda}}
&3\lambda\gg32\pi^2 m^4/H^4
\end{cases}
\end{equation}
The comparison of these two expressions leads to the idea of an  `effective mass'
of order $\lambda^{1/4}H$ in the $m\to0$ limit \cite{Garbrecht:2011gu, Beneke:2012kn}. However, this does not play the role of a true mass in the following sense: if we look at
the `$00$' component  stress-energy tensor we obtain
\begin{equation}
\langle T_{00}\rangle_{\rm eq}\sim
\begin{cases}
\displaystyle{3H^4\over 32\pi^2}&3\lambda\ll32\pi^2 m^4/H^4\\[2ex]
\displaystyle{\sqrt{3}\Gamma(3/4)^2\over 8\pi^2}{H^2m^2\over \sqrt{\lambda}}
&3\lambda\gg32\pi^2 m^4/H^4
\end{cases}
\end{equation}
In the $m\to0$ limit $\langle T_{00}\rangle_{\rm eq}\to0$, whereas a free scalar field with a mass 
$m\sim\lambda^{1/4}H$ would give a quite different result 
$\langle T_{00}\rangle_{\rm eq}\sim H^4$. The above expression also shows that IR fluctuations do not contribute to the energy density of a massless self interacting scalar field; all such contributions come from UV fluctuations, see figure 2.

\section{Discussion}
We presented a detailed derivation of the effective IR theory of a light scalar field $\phi$ in de Sitter spacetime. By explicitly integrating out subhorizon modes, we obtained the stochastic dynamics first discussed by Starobinsky, as well as the next-to-leading-order corrections to the original stochastic picture. A new feature is that the dynamics is second order in time, meaning that both $\phi$ and its velocity are included, with the latter equilibrating to a Maxwell distribution on a timescale $\Delta t \sim 1/H$. The origin of the dominant stochastic term is the time dependence of the split between long and short modes. Including self interactions that couple long and short modes results in the long wavelength field evolving on a shifted effective potential whose renormalized form we compute. Self-interactions also result in subdominant stochastic forces of the kind encountered when there is a system-environment coupling. We hope our derivation further clarifies some existent misunderstandings on the origin of Starobinsky's noise term which is present even if there is no self-coupling. 

We also derived a functional Kramers equation for the long wavelegth Wigner functional which can be used to compute spacetime expectation values of fields and field velocities. When restricted to coincident expectation values, the equation can be simplified to an ordinary Kramers equation, admitting an equilibrium solution that takes the Maxwell - Boltzmann form for the field averaged over a spatial volume $\sim \frac{4\pi}{3}H^{-3}$ and at a temperature $T=H/2\pi$ and an effective potential resulting from integrating out the UV. 

If the IR action (\ref{MSRJD-2}) is used to define Feynman rules, individual diagrams will exhibit the same IR divergences as the full QFT diagrams. However, the stochastic theory can also be formulated in terms of a Wigner functional providing a probability distribution wrt which IR finite expectation values can be computed. It will be interesting to see if the methods developed here will be applicable to more complicated theories, including gravitational fluctuations, in the hope of resolving the IR issues they exhibit. It may also be of interest to use the dynamical stochastic formalism to investigate symmetry restoration in de Sitter studied in \cite{Lazzari:2013boa, Guilleux:2015pma, Gonzalez:2016jrn, Guilleux:2016oqv}.

\acknowledgments
IGM is supported by STFC (Consolidated Grant ST/J000426/1). The authors would like to thank the anonymous referee for their thorough comments on an earlier version of the manuscript. GR would also like to thank T. Prokopec for very useful discussions on the topics of this paper.      

\appendix

\section{Deriving the effective action in the Keldysh Basis}
We present here the derivation of the effective action, unpacking the index notation of section \ref{eff IR act} in the Keldysh basis. For our computations it suffices to only consider terms linear in $\phi_{\rm q}$ in $\overline{V}_I$ (\ref{Vbar})
\beq\label{Vbar approx}
\overline{V}_I\simeq V_I'\phi_{\rm q}\,,
\eeq   
and 
\bea
\mathbf{D}_>^{-1ab}+{\delta S_I\over\delta\phi_{<a}\delta\phi_{<b}}&\simeq& \left(\begin{matrix}
	0& \nabla^2-\tilde{m}^2 \\ \nabla^2-\tilde{m}^2 & 0	
\end{matrix}\right)\delta(x-x') + \left(\begin{matrix}  {\partial^3 V_I\over\partial\phi_<{}^3}\phi_{\rm q <}
& 0 \\ 0 & 0	
\end{matrix}\right)\delta(x-x')\\
&\equiv& \tilde{\mathbf{D}}_>^{-1ab} + \mathbf{V}_3^{ab}\,,
\eea
where
\beq\label{tildem}
\tilde{m}^2(\phi_<)=m^2 - V''(\phi_{<})\,.
\eeq
We therefore have 
\beq
-\frac{i}{2}\ln \det (\mathbf{G}_>) \simeq \frac{i}{2}{\rm Tr} \ln \left(1+\tilde{\mathbf{D}}_>\mathbf{V}_3  \right) 
\eeq
since ${\rm Tr}\ln \tilde{\mathbf{D}}_>=0$, see (\ref{Z0}, \ref{Z0-2}), and we obtain
\begin{equation}\label{Veff1}
-\frac{i}{2}\ln\det (\mathbf{G}_>)\simeq \frac{1}{2}\int d\mu(x)D^>_{K}(x,x)
{\partial^3 V_I\over\partial\phi_<{}^3}\phi_{\rm q <}(x).
\end{equation}
The coincident limit of the $D_{K}^>$ propagator corresponds to the same limit of the more familiar Feynman propagator and hence harbours the same UV divergence. 

The contribution (\ref{Veff1}) to the effective action adds a correction $\Delta V$ to the interacting potential experienced by the $\phi_<$ field
\beq
V_{I \rm eff}(\phi_<) = V_I(\phi_<) + \Delta V (\phi_<)\,,
\eeq
and this enters the effective Keldysh action through 
\beq
\overline{V}_{I\rm eff}(\phi_<) \simeq \frac{\partial V_{I\rm eff}}{\partial \phi_<}\phi_{\rm q <}\,.
\eeq  
$\Delta V$ must therefore satisfy
\beq\label{eqVeff}
\frac{\partial \Delta V}{\partial\phi_<} = \frac{1}{2} D^>_{K}(x,x)
{\partial^3 V_I\over\partial\phi_<{}^3}\,.
\eeq
Note that $D_{K}^>$ also depends on $\phi_<$ through (\ref{tildem}).  

The last term in (\ref{Seff1}) adds to the effective action 
\beq\label{DeltaS2}
\Delta S_2 = -\int \psi_< V_I''D^>_RV' + \frac{i}{2}
\int  \psi_< V_I'' D^>_K V_I'' \psi_<\,,
\eeq 
where integration over the two arguments of $D$ is implied. The second term in $\Delta S_2$ corresponds to a stochastic force $f$ on the long wavelength modes with a coloured correlation function
\beq\label{extra stoch}
\langle f(x)f(x')\rangle = V_I''(x)D_k(x,x')V_I''(x')\,.
\eeq
This force acts in addition to the Starobinsky noise term discussed in the main text and results from the short wavelength fluctuations due to the self coupling of the field. The first term in (\ref{DeltaS2}) adds an a priori non-local force term to the equations of motion. However, by restricting $D_R^>$ to timescales not significantly longer than $\Delta t\sim H^{-1}$ we have $D_R^>\sim \frac{-1}{3H} e^{-3H(t-t')}\Theta(t-t')\delta(\vc{x}-\vc{x}')$ and hence we obtain an addition to the effective potential of $\phi_<$ of the form 
\beq
\Delta V \approx -\int V_I'\frac{V_I''}{9H^2} d\phi\,,
\eeq 
which is second order in the non-linear coupling.

\section{Long time/strong friction limit and the Fokker-Planck equation} 
The more familiar Fokker-Planck equation can be obtained by standard analysis \cite{Risken} when considering timescales $\Delta t \gg H^{-1}$. Reference \cite{Riotto:2011sf} provides a detailed exposition of this procedure in cosmologically relevant examples. A simple approach amounts to setting $H^{-1}\rightarrow 0$. Using $1/3H$ as an expansion parameter it is possible to obtain 
\beq
W(\phi,v,t) = \left(\overline{W}_0(\phi) - \frac{v}{3H}\left(\frac{\partial \overline{W}_0}{\partial\phi} +
\frac{d U}{d\phi}\Omega \overline{W}_0\right) + \overline{W}_1(\phi, t)\frac{1}{3H} +
\mc{O}\left(\frac{1}{9H^2}\right)\right)e^{-\Omega v^2/2}\,,
\eeq
where
\beq
\frac{\partial}{\partial t} \overline{W}_1 = \frac{1}{\Omega}\frac{\partial}{\partial\phi}\left(\frac{\partial}{\partial\phi} + \Omega\frac{d V}{d\phi}\right)\overline{W}_0\,.
\eeq
Integrating over $v$ we see that, to $\mathcal{O}(1/3H)$,
\beq\label{FP3}
\frac{\partial P}{\partial t}= \frac{1}{3H\Omega} \frac{\partial}{\partial\phi}\left(\frac{\partial}{\partial\phi}+
\Omega\frac{d V}{d\phi}\right)P\,,
\eeq
where $P(\phi,t)=\int dv W(\phi,v,t)$. Equation (\ref{FP3}) is the more familiar Starobinsky-Fokker-Planck equation for $\phi$. Hence Starobinsky's original stochastic equation can be used when processes with time scales $\Delta t \sim H^{-1}$ are neglected.


\begin{thebibliography}{199}

\bibitem{Ford:1977in}
L.~H.~Ford and L.~Parker,
``Infrared Divergences in a Class of Robertson-Walker Universes,''
Phys.\ Rev.\ D {\bf 16} (1977) 245

\bibitem{Seery:2010kh}
D.~Seery,
``Infrared effects in inflationary correlation functions,''
Class.\ Quant.\ Grav.\  {\bf 27} (2010) 124005
[arXiv:1005.1649 [astro-ph.CO]]

\bibitem{Woodard:2014jba}
R.~P.~Woodard,
``Perturbative Quantum Gravity Comes of Age,''
Int.\ J.\ Mod.\ Phys.\ D {\bf 23} (2014) no.09,  1430020
[arXiv:1407.4748 [gr-qc]].

\bibitem{Starobinsky:1986fx}
A.~A.~Starobinsky,
``Stochastic De Sitter (inflationary) Stage In The Early Universe,''
In *De Vega, H.j. ( Ed.), Sanchez, N. ( Ed.): Field Theory, Quantum Gravity and Strings*, 107-126

\bibitem{Starobinsky:1994bd}
A.~A.~Starobinsky and J.~Yokoyama,
``Equilibrium state of a selfinteracting scalar field in the De Sitter background,''
Phys.\ Rev.\ D {\bf 50} (1994) 6357
[astro-ph/9407016].

\bibitem{Woodard:2005cw}
R.~P.~Woodard,
``A Leading logarithm approximation for inflationary quantum field theory,''
Nucl.\ Phys.\ Proc.\ Suppl.\  {\bf 148} (2005) 108
[astro-ph/0502556].

\bibitem{vanderMeulen:2007ah}
M.~van der Meulen and J.~Smit,
``Classical approximation to quantum cosmological correlations,''
JCAP {\bf 0711} (2007) 023
[arXiv:0707.0842 [hep-th]].

\bibitem{Finelli:2008zg}
F.~Finelli, G.~Marozzi, A.~A.~Starobinsky, G.~P.~Vacca and G.~Venturi,
``Generation of fluctuations during inflation: Comparison of stochastic and field-theoretic approaches,''
Phys.\ Rev.\ D {\bf 79} (2009) 044007
[arXiv:0808.1786 [hep-th]].

\bibitem{Garbrecht:2014dca}
B.~Garbrecht, F.~Gautier, G.~Rigopoulos and Y.~Zhu,
``Feynman Diagrams for Stochastic Inflation and Quantum Field Theory in de Sitter Space,''
Phys.\ Rev.\ D {\bf 91} (2015) 063520
[arXiv:1412.4893 [hep-th]].

\bibitem{Garbrecht:2013coa}
B.~Garbrecht, G.~Rigopoulos and Y.~Zhu,
``Infrared correlations in de Sitter space: Field theoretic versus stochastic approach,''
Phys.\ Rev.\ D {\bf 89} (2014) 063506
[arXiv:1310.0367 [hep-th]].

\bibitem{Onemli:2015pma}
V.~K.~Onemli,
``Vacuum Fluctuations of a Scalar Field during Inflation: Quantum versus Stochastic Analysis,''
Phys.\ Rev.\ D {\bf 91} (2015) 103537
[arXiv:1501.05852 [gr-qc]].

\bibitem{Rajaraman:2010xd}
A.~Rajaraman,
Phys.\ Rev.\ D {\bf 82} (2010) 123522
doi:10.1103/PhysRevD.82.123522
[arXiv:1008.1271 [hep-th]].

\bibitem{Beneke:2012kn}
M.~Beneke and P.~Moch,
``On “dynamical mass” generation in Euclidean de Sitter space,''
Phys.\ Rev.\ D {\bf 87} (2013) 064018
doi:10.1103/PhysRevD.87.064018
[arXiv:1212.3058 [hep-th]].

\bibitem{Guilleux:2015pma}
M.~Guilleux and J.~Serreau,
``Quantum scalar fields in de Sitter space from the nonperturbative renormalization group,''
Phys.\ Rev.\ D {\bf 92} (2015) no.8,  084010
[arXiv:1506.06183 [hep-th]].

\bibitem{Gautier:2015pca}
F.~Gautier and J.~Serreau,
``Scalar field correlator in de Sitter space at next-to-leading order in a 1/N expansion,''
Phys.\ Rev.\ D {\bf 92} (2015) no.10,  105035
[arXiv:1509.05546 [hep-th]].

\bibitem{Nacir:2016fzi}
D.~L$\acute{o}$pez Nacir, F.~D.~Mazzitelli and L.~G.~Trombetta,
JHEP {\bf 1609} (2016) 117
doi:10.1007/JHEP09(2016)117
[arXiv:1606.03481 [hep-th]].

\bibitem{Burgess:2015ajz}
C.~P.~Burgess, R.~Holman and G.~Tasinato,
``Open EFTs, IR effects and late-time resummations: systematic corrections in stochastic inflation,''
JHEP {\bf 1601} (2016) 153
[arXiv:1512.00169 [gr-qc]].

\bibitem{Nambu:1988je}
Y.~Nambu and M.~Sasaki,
``Stochastic Approach to Chaotic Inflation and the Distribution of Universes,''
Phys.\ Lett.\ B {\bf 219} (1989) 240.

\bibitem{Kandrup:1988sc}
H.~E.~Kandrup,
``Stochastic Inflation As A Time Dependent Random Walk,''
Phys.\ Rev.\ D {\bf 39} (1989) 2245.

\bibitem{Salopek:1990re}
D.~S.~Salopek and J.~R.~Bond,
``Stochastic inflation and nonlinear gravity,''
Phys.\ Rev.\ D {\bf 43} (1991) 1005.

\bibitem{Mollerach:1990zf}
S.~Mollerach, S.~Matarrese, A.~Ortolan and F.~Lucchin,
``Stochastic inflation in a simple two field model,''
Phys.\ Rev.\ D {\bf 44} (1991) 1670.

\bibitem{Habib:1992ci}
S.~Habib,
``Stochastic inflation: The Quantum phase space approach,''
Phys.\ Rev.\ D {\bf 46} (1992) 2408
[gr-qc/9208006].

\bibitem{GarciaBellido:1994vz}
J.~Garcia-Bellido,
``Jordan-Brans-Dicke stochastic inflation,''
Nucl.\ Phys.\ B {\bf 423} (1994) 221
[astro-ph/9401042].

\bibitem{Liguori:2004fa}
M.~Liguori, S.~Matarrese, M.~Musso and A.~Riotto,
``Stochastic inflation and the lower multipoles in the CMB anisotropies,''
JCAP {\bf 0408} (2004) 011
[astro-ph/0405544].

\bibitem{Rigopoulos:2004gr}
G.~I.~Rigopoulos and E.~P.~S.~Shellard,
``Non-linear inflationary perturbations,''
JCAP {\bf 0510} (2005) 006
[astro-ph/0405185].

\bibitem{Martin:2005ir}
J.~Martin and M.~Musso,
``Solving stochastic inflation for arbitrary potentials,''
[hep-th/0511214].

\bibitem{Prokopec:2007ak}
T.~Prokopec, N.~C.~Tsamis and R.~P.~Woodard,
``Stochastic Inflationary Scalar Electrodynamics,''
Annals Phys.\  {\bf 323} (2008) 1324
[arXiv:0707.0847 [gr-qc]].

\bibitem{Adshead:2008gk}
P.~Adshead, R.~Easther and E.~A.~Lim,
``Cosmology With Many Light Scalar Fields: Stochastic Inflation and Loop Corrections,''
Phys.\ Rev.\ D {\bf 79} (2009) 063504
[arXiv:0809.4008 [hep-th]].

\bibitem{Riotto:2008mv}
A.~Riotto and M.~S.~Sloth,
``On Resumming Inflationary Perturbations beyond One-loop,''
JCAP {\bf 0804} (2008) 030
[arXiv:0801.1845 [hep-ph]].

\bibitem{Finelli:2010sh}
F.~Finelli, G.~Marozzi, A.~A.~Starobinsky, G.~P.~Vacca and G.~Venturi,
``Stochastic growth of quantum fluctuations during slow-roll inflation,''
Phys.\ Rev.\ D {\bf 82} (2010) 064020
[arXiv:1003.1327 [hep-th]].

\bibitem{Kuhnel:2010pp}
F.~Kuhnel and D.~J.~Schwarz,
``Large-Scale Suppression from Stochastic Inflation,''
Phys.\ Rev.\ Lett.\  {\bf 105} (2010) 211302
[arXiv:1003.3014 [hep-ph]].

\bibitem{Riotto:2011sf}
A.~Riotto and M.~S.~Sloth,
``The probability equation for the cosmological comoving curvature perturbation,''
JCAP {\bf 1110} (2011) 003
[arXiv:1103.5876 [astro-ph.CO]].

\bibitem{Enqvist:2011pt}
K.~Enqvist, D.~G.~Figueroa and G.~Rigopoulos,
``Fluctuations along supersymmetric flat directions during Inflation,''
JCAP {\bf 1201} (2012) 053
[arXiv:1109.3024 [astro-ph.CO]].

\bibitem{Weenink:2011dd}
J.~Weenink and T.~Prokopec,
``On decoherence of cosmological perturbations and stochastic inflation,''
arXiv:1108.3994 [gr-qc].

\bibitem{Martin:2011ib}
J.~Martin and V.~Vennin,
``Stochastic Effects in Hybrid Inflation,''
Phys.\ Rev.\ D {\bf 85} (2012) 043525
[arXiv:1110.2070 [astro-ph.CO]].

\bibitem{Hwang:2012mf}
D.~i.~Hwang, B.~H.~Lee, E.~D.~Stewart, D.~h.~Yeom and H.~Zoe,
``Euclidean quantum gravity and stochastic inflation,''
Phys.\ Rev.\ D {\bf 87} (2013) no.6,  063502
[arXiv:1208.6563 [gr-qc]].

\bibitem{Rigopoulos:2013exa}
G.~Rigopoulos,
``Fluctuation-dissipation and equilibrium for scalar fields in de Sitter,''
arXiv:1305.0229 [astro-ph.CO].

\bibitem{Levasseur:2013tja}
L.~Perreault Levasseur, V.~Vennin and R.~Brandenberger,
``Recursive Stochastic Effects in Valley Hybrid Inflation,''
Phys.\ Rev.\ D {\bf 88} (2013) 083538
[arXiv:1307.2575 [hep-th]].

\bibitem{Lazzari:2013boa}
G.~Lazzari and T.~Prokopec,
``Symmetry breaking in de Sitter: a stochastic effective theory approach,''
arXiv:1304.0404 [hep-th].

\bibitem{Burgess:2014eoa}
C.~P.~Burgess, R.~Holman, G.~Tasinato and M.~Williams,
``EFT Beyond the Horizon: Stochastic Inflation and How Primordial Quantum Fluctuations Go Classical,''
JHEP {\bf 1503} (2015) 090
[arXiv:1408.5002 [hep-th]].

\bibitem{Vennin:2015hra}
V.~Vennin and A.~A.~Starobinsky,
``Correlation Functions in Stochastic Inflation,''
Eur.\ Phys.\ J.\ C {\bf 75} (2015) 413
[arXiv:1506.04732 [hep-th]].

\bibitem{Vennin:2016wnk}
V.~Vennin, H.~Assadullahi, H.~Firouzjahi, M.~Noorbala and D.~Wands,
``Critical Number of Fields in Stochastic Inflation,''
arXiv:1604.06017 [astro-ph.CO].

\bibitem{Boyanovsky:2015tba}
D.~Boyanovsky,
``Effective field theory during inflation: Reduced density matrix and its quantum master equation,''
Phys.\ Rev.\ D {\bf 92} (2015) no.2,  023527
[arXiv:1506.07395 [astro-ph.CO]].

\bibitem{Boyanovsky:2015jen}
D.~Boyanovsky,
``Effective field theory during inflation. II. Stochastic dynamics and power spectrum suppression,''
Phys.\ Rev.\ D {\bf 93} (2016) 043501
[arXiv:1511.06649 [astro-ph.CO]].


\bibitem{Rigopoulos:2016oko}
G.~Rigopoulos,
``Thermal Interpretation of Infrared Dynamics in de Sitter,''
JCAP {\bf 1607} (2016) no.07,  035
[arXiv:1604.04313 [gr-qc]].

\bibitem{Morikawa:1989xz}
M.~Morikawa,
``Dissipation And Fluctuation Of Quantum Fields In Expanding Universes,''
Phys.\ Rev.\ D {\bf 42} (1990) 1027.

\bibitem{Levasseur:2013ffa}
L.~Perreault Levasseur,
``Lagrangian formulation of stochastic inflation: Langevin equations, one-loop corrections and a proposed recursive approach,''
Phys.\ Rev.\ D {\bf 88} (2013) no.8,  083537
[arXiv:1304.6408 [hep-th]].

\bibitem{Tsamis:2005hd}
N.~C.~Tsamis and R.~P.~Woodard,
``Stochastic quantum gravitational inflation,''
Nucl.\ Phys.\ B {\bf 724} (2005) 295
[gr-qc/0505115].

\bibitem{Calzetta:2008iqa}
E.~A.~Calzetta and B.~L.~B.~Hu,
``Nonequilibrium Quantum Field Theory,''
Cambridge, UK: Univ. Pr. (2008) 

\bibitem{Altland:2006si}
A.~Altland and B.~Simons,
``Condensed matter field theory,''
Cambridge, UK: Univ. Pr. (2010), 2nd ed

\bibitem{Delamotte:2007pf}
B.~Delamotte,
``An Introduction to the nonperturbative renormalization group,''
Lect.\ Notes Phys.\  {\bf 852} (2012) 49
[cond-mat/0702365 [cond-mat.stat-mech]].

\bibitem{ZinnJustin:2002ru}
J.~Zinn-Justin,
``Quantum field theory and critical phenomena,''
Int.\ Ser.\ Monogr.\ Phys.\  {\bf 113} (2002) 1.

\bibitem{Chernikov:1968zm}
  N.~A.~Chernikov and E.~A.~Tagirov,
  Ann.\ Inst.\ H.\ Poincare Phys.\ Theor.\ A {\bf 9} (1968) 109.

\bibitem{Bunch:1978yq}
  T.~S.~Bunch and P.~C.~W.~Davies,
  Proc.\ Roy.\ Soc.\ Lond.\ A {\bf 360} (1978) 117.
  doi:10.1098/rspa.1978.0060



\bibitem{Polarski:1995jg}
D.~Polarski and A.~A.~Starobinsky,
``Semiclassicality and decoherence of cosmological perturbations,''
Class.\ Quant.\ Grav.\  {\bf 13} (1996) 377
[gr-qc/9504030].

\bibitem{Risken}
H.~Risken
``The Fokker Planck Equation: Methods of Solution and Applications'', Springer, 2nd ed

\bibitem{Graziani:1988}
F.~R.~Graziani
"Quantum probability distributions in the early Universe. I. Equilibrium properties of the Wigner equation",
Phys.\ Rev.\ D {\bf 38} (1988) 1122

\bibitem{Buryak:1995tx}
O.~E.~Buryak,
``Stochastic dynamics of large scale inflation in de Sitter space,''
Phys.\ Rev.\ D {\bf 53} (1996) 1763
[gr-qc/9502032].

\bibitem{Vilenkin:1982wt}
A.~Vilenkin and L.~H.~Ford,
``Gravitational Effects upon Cosmological Phase Transitions,''
Phys.\ Rev.\ D {\bf 26} (1982) 1231.

\bibitem{Linde:1982uu}
A.~D.~Linde,
``Scalar Field Fluctuations in Expanding Universe and the New Inflationary Universe Scenario,''
Phys.\ Lett.\  {\bf 116B} (1982) 335.

\bibitem{Starobinsky:1982ee}
A.~A.~Starobinsky,
``Dynamics of Phase Transition in the New Inflationary Universe Scenario and Generation of Perturbations,''
Phys.\ Lett.\  {\bf 117B} (1982) 175.

\bibitem{Vilenkin:1983xp}
A.~Vilenkin,
``Quantum Fluctuations in the New Inflationary Universe,''
Nucl.\ Phys.\ B {\bf 226} (1983) 527.

\bibitem{Clifton:2014fja}
T.~Clifton and J.~D.~Barrow,
Fundam.\ Theor.\ Phys.\  {\bf 187} (2017) 61
[arXiv:1412.5465 [gr-qc]].

\bibitem{Markkanen:2016jhg}
T.~Markkanen,
``Decoherence Can Relax Cosmic Acceleration,''
JCAP {\bf 1611} (2016) no.11,  026
[arXiv:1609.01738 [hep-th]].

\bibitem{Garbrecht:2011gu}
B.~Garbrecht and G.~Rigopoulos,
``Self Regulation of Infrared Correlations for Massless Scalar Fields during Inflation,''
Phys.\ Rev.\ D {\bf 84} (2011) 063516
[arXiv:1105.0418 [hep-th]].

\bibitem{Birrell:1982ix}
N.~D.~Birrell and P.~C.~W.~Davies,
``Quantum Fields In Curved Space,''
Cambridge, Uk: Univ. Pr. (1982) 

\bibitem{Gonzalez:2016jrn}
  F.~F.~Gonzalez and T.~Prokopec,
  arXiv:1611.07854 [gr-qc].

\bibitem{Guilleux:2016oqv}
  M.~Guilleux and J.~Serreau,
  Phys.\ Rev.\ D {\bf 95} (2017) no.4,  045003
  doi:10.1103/PhysRevD.95.045003
  [arXiv:1611.08106 [gr-qc]].

\end{thebibliography}
\end{document}